\documentclass[%
 reprint,
 amsmath,amssymb,
 aps,
prb,
]{revtex4-2}
\usepackage{graphicx} 
\usepackage{comment}
\usepackage{svg}
\usepackage[breaklinks=true,colorlinks,citecolor=blue,linkcolor=blue,urlcolor=blue]{hyperref}
\usepackage{cleveref}

\newcommand{\imag}{\text{i}}
\newcommand{\diff}{\text{d}}
\newcommand{\mub}{\,\mu_\text{B}}

\usepackage[normalem]{ulem}
\newcommand{\red}[1]{\begingroup\color{black}#1\endgroup}
\newcommand{\blue}[1]{\begingroup\color{black}#1\endgroup}
\renewcommand{\sout}[1]{}

\crefname{equation}{Eq.}{Eqs.}
\Crefname{equation}{Equation}{Equations}
\crefname{figure}{Fig.}{Figs.}
\Crefname{figure}{Figure}{Figures}
\crefname{table}{Table}{Tables}
\Crefname{table}{Table}{Tables}

\begin{document}

\title{Capturing \red{exchange-correlation} spin-torque effects with a semilocal functional}

\author{Marie-Therese Huebsch}
\email{marie-therese.huebsch@vasp.at}
\affiliation{VASP Software GmbH, Berggasse 21/14, A-1090, Vienna, Austria}

\author{Fabien Tran}%
\affiliation{VASP Software GmbH, Berggasse 21/14, A-1090, Vienna, Austria}

\author{Martijn Marsman}%
\affiliation{VASP Software GmbH, Berggasse 21/14, A-1090, Vienna, Austria}

\date{\today}

\begin{abstract}

    We cure the lack of \red{exchange-correlation (XC)} spin torque in semilocal \sout{exchange-correlation (XC)} \red{XC} functionals by treating XC effects in the framework of spin-current-density-functional theory (SCDFT), and present the implementation of the first kind of this novel family of XC functionals in the Vienna ab-initio simulation package (VASP): An U(1)$\times$SU(2) gauge-invariant SCDFT functional featuring a $2\times 2$ XC potential. While the framework can be applied to other XC functionals, the presented flavor of the SCDFT functional is based on Becke-Roussel exchange and Colle-Salvetti correlation. In addition to the $2\times 2$ spin density and kinetic-energy density, the XC functional depends on the $2\times 2$ spin-current density. The implementation requires the computation of the spin-current density within the projector-augmented-wave method and the variation of the XC energy with respect to it. The application to a Cr$_3$ molecule and bulk MnO reveals (i) \red{XC} spin torque of the same order as obtained by methods including exact exchange, (ii) a counterintuitive contribution to the energy even in collinear ferromagnetic systems without spin-orbit coupling due to the semilocality of the magnetization, and (iii) a similar computational cost per electronic step as calculations that depend on, inter alia, the kinetic-energy density, but convergence within fewer electronic steps. 
\end{abstract}

\maketitle

\section{Introduction}
\label{sec:Introduction}

Magnetic ab-initio calculations using semilocal exchange-correlation (XC) functionals suffer from a fundamental flaw:  Their ignorance of \red{XC} spin torque. This is possibly relevant for systems featuring noncollinear magnetism, disordered magnetic moments, spin dynamics,  magnetic frustration, spin waves, skyrmions, and systems with static currents such as in the presence of orbital moments, spin Hall effect, in nuclear magnetic resonance, etc. Curiously, even collinear magnetic systems may observe \red{XC} spin torque and, hence, already show this deficiency. 

\begin{figure}[b]
\includegraphics[width=0.7\columnwidth]{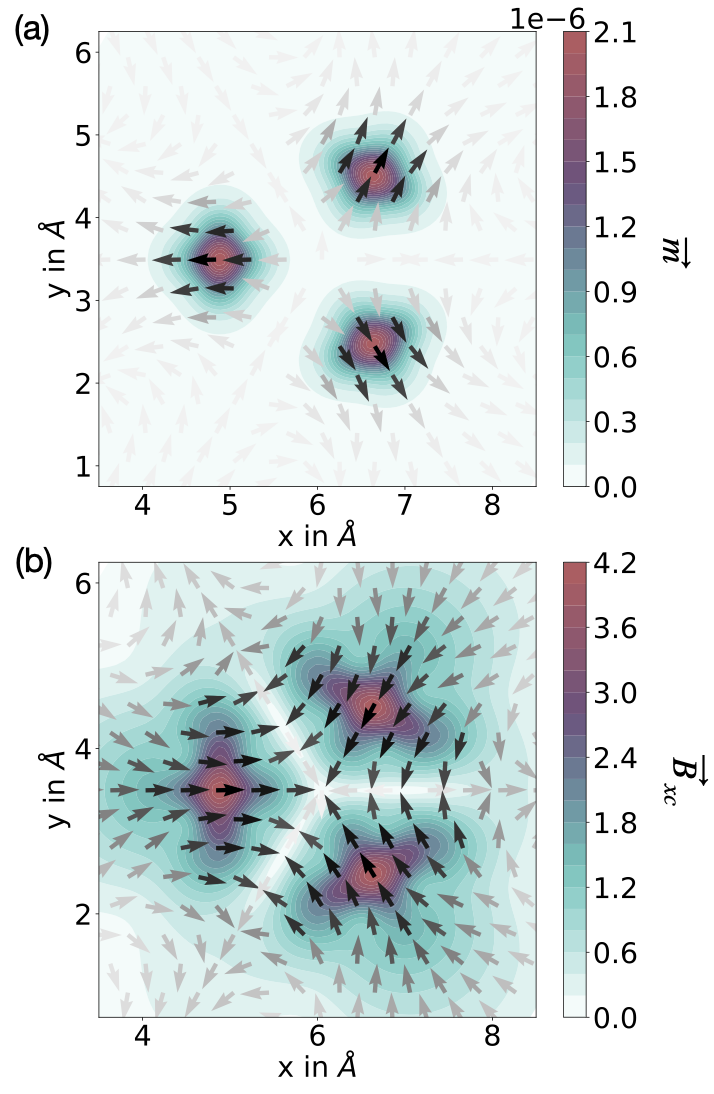}
\caption{(a) Magnetization $\vec{m}$ and (b) XC magnetic B-field $\vec{B}_\text{xc}$ in Cr$_{3}$ obtained from a LSDA calculation. Computational details are discussed in \Cref{subsec:Isolated Cr3}. \label{fig1:Cr trimer}}
\end{figure}

Measured by the amount of magnetic ab-initio calculations performed, the issue has been discussed relatively peripherally. To provide a balanced perspective, when Kübler \emph{et al.} \cite{kubler1988local,kubler1988density} proposed an algorithm to treat noncollinear magnetic systems from first principles, they already pointed out that the transformation to a local spin-quantization axis only holds for the local spin-density approximation (LSDA) \cite{Kohn1965}. In other words, within LSDA the XC magnetic B-field $B_{\text{xc},a}=\delta E_{\text{xc}}/\delta m_{a}$ aligns with the magnetization $\vec{m}$ at each point $\mathbf{r}$ in real space and the system is therefore spin-torque-free: $\vec{m}\times\vec{B}_{\text{xc}}=0$ for LSDA. This procedure allows determining the direction of the magnetization self-consistently within density-functional theory (DFT) \cite{Hohenberg1964} while  evaluating the XC functional for local spin-up $n_\uparrow$ and spin-down $n_\downarrow$ densities, i.e., using a mere collinear, spin-polarized XC functional $E_{\text{xc}}[n_\uparrow,n_\downarrow]$ for noncollinear calculations. \Cref{fig1:Cr trimer}(a)-(b) illustrate how the magnetization and XC B-field of an isolated Cr trimer in 
a traditional noncollinear calculation 
are absolutely parallel, yielding zero \red{XC} spin torque. 

The DFT community got used to employing XC functionals beyond the LSDA for magnetic calculations. For instance, XC functionals of the generalized gradient approximation (GGA), meta-GGA including the kinetic-energy density, or more rarely hybrid functionals, that include a fraction of Hartree-Fock exchange \cite{PerdewAIP01}.
In fact, many implementations of noncollinear DFT
allow the use of any collinear XC functional despite the missing theoretical foundation to use functionals beyond the LSDA for noncollinear calculations.
This is particularly problematic if the semilocal nature, i.e., gradient corrections, significantly influences the electronic structure and spin texture.
The lack of \red{XC} spin torque was criticized quickly
by Kleinman 
\cite{kleinman1999density}, 
but the first attempts at formulating a noncollinear XC functional by introducing oﬀ-diagonal elements to the XC potential led to disappointing numerical results \cite{bylander1999full}. 

The theoretical understanding of this failure is given in a series of works around the year 2000, where Capelle \emph{et al.}~\cite{capelle1997spin,capelle2000density,capelle2001spin} analyze time-dependent spin-density-functional theory (SDFT) and formulate the so-called zero-torque theorem: It states that the XC B-field cannot exert a net global spin torque on the system. Perhaps even more interesting, in the same works, they established a clear \emph{connection between the \red{XC} spin torque and the spin-current density}. They show that in the static limit with no external B-field and \emph{without} spin-orbit coupling (SOC), the divergence of the tensor-valued XC spin-current density $\mathbf{\nabla} \cdot \vec{\mathbf{J}}_{\text{xc}}$---i.e., the difference between the many-body and Kohn-Sham (KS) currents \cite{capelle2001spin}---generates a local \red{XC} spin torque $\vec{m} \times \vec{B}_{\text{xc}}$. This renders any spin-torque-free XC functional to be at least an approximation when studying systems governed by spin-dynamic effects, e.g., any itinerant magnet apart from the fully spin-polarized homogeneous electron gas. This shows that the spin-current density, and thus \blue{spin-current-density-functional theory} (SCDFT) \cite{vignale1988current}, should play an integral role when extending the state-of-the-art framework of noncollinear ab-initio calculations.

It is worthwhile at this point to mention the most recent works reporting the implementation of functionals depending on the current density, even if they do not necessarily focus on magnetism. Sen and Tellgren proposed a new family of functionals beyond the meta-GGA that depend also on the paramagnetic vorticity \cite{SenJCP2018}. For a correct treatment of SOC, Trushin and G\"{o}rling reported a fully self-consistent implementation of a general unifying theory using exact exchange (EXX) \cite{TrushinPRB2018}. For the geometry optimization of molecules in strong magnetic fields, Irons, David, and Teale \cite{Irons2021} implemented MGGA and hybrid functionals that also depend on the current density. 
Holzer, Franzke, and Pausch \cite{holzer2022current,FranzkeJCP2024}, as well as Richter \emph{et al.} \cite{RichterJCP2023}, applied existing meta-GGA functionals with a kinetic-energy density that is augmented by a term depending on the paramagnetic current density. 

Returning to our main focus, curing the lack of \red{XC} spin torque: A suitable guide towards proper noncollinear XC functionals are methods that map the nonlocal Hartree-Fock-exchange potential onto a local EXX potential, i.e., calculations using the optimized effective potential (OEP) method \cite{Sharp1953:317,Talman1976:36}. However, EXX calculations come at high computational cost, rendering them unfeasible for realistic applications, yet they offer a glance at the main patterns of \red{XC} spin torque \cite{sharma2007first,tancogne2023constructing,tancogne2023constructing_erratum}. 
Specifically, the \red{XC} spin torque of the same isolated Cr trimer as shown in \cref{fig1:Cr trimer} is shown in Fig.~1(a) and (b) of Ref.~\cite{tancogne2023constructing} using (a) the Slater potential \cite{SlaterPR1951,UllrichPRB2018}, which is a component of the EXX potential, and (b) the more accurate Krieger-Li-Iafrate (KLI) approximation to the EXX-OEP method \cite{krieger1992construction}.

In that light, thus far, approaches to formulate a noncollinear XC functional \cite{kleinman1999density,capelle2000density,sjostedt2002noncollinear,katsnelson2003spin,capelle2003exploring,bencheikh2003spin,peralta2007noncollinear,scalmani2012new,bulik2013noncollinear,eich2013transverse,sharma2018source,pu2023noncollinear,tancogne2023constructing,tancogne2023constructing_erratum} have not proven satisfactory either because they do not obey the zero-torque theorem, neglect the spin-current density, do not reproduce the \red{XC} spin torque, or break the U(1)$\times$SU(2) invariance.
In an effort to overcome this limitation, we follow in this work an approach that is substantially inspired by Pittalis, Vignale, and Eich \cite{pittalis2017u} and augmented by the correlation-energy functional developed by Tancogne-Dejean, Rubio, and Ullrich \cite{tancogne2023constructing,tancogne2023constructing_erratum}. 
The lack of \red{XC} spin torque is cured by introducing a semilocal XC functional within \sout{spin-current-density-functional theory} SCDFT. The complete SCDFT functional depends on the $2\times2$ spin density, the $2\times2$ kinetic-energy density, and the three vector components of the $2\times2$ paramagnetic spin-current density. 
This yields associated $2\times2$ partial derivatives in the expression of the $2\times2$ XC potential. \red{Mind that in the present work, we discuss the implementation and results on noncollinear magnetism and leave the discussion of SOC effects in the context of SCDFT for prospective publications.}

To the best of our knowledge, we report the first full \red{projector-augmented-wave (PAW)} implementation of a static noncollinear SCDFT functional, where the SDFT variational problem is extended by the variation with respect to the components of the $2\times2$ paramagnetic spin-current density and the XC functional explicitly provides a $2\times2$ potential. 
Intriguingly, the derivation by Pittalis, Vignale, and Eich \cite{pittalis2017u} should offer a formal way of generalizing other existing functionals to noncollinear functionals yielding a non-zero local \red{XC} spin torque. Promising recent works on SCDFT and U(1)$\times$SU(2) symmetry by Desmarais \emph{et al.} \cite{desmarais2020adiabatic,DesmaraisPRM2024,desmarais2024electron,DesmaraisPRL2024} 
pave the way to also generalizing other XC functionals \sout{such as the strongly constrained and appropriately normed (SCAN) functionals [42-44]}  to be used within SCDFT. 
\red{In fact, shortly after submitting our manuscript, we learned about a parallel work by Desmarais, Erba, Vignale and Pittalis \cite{desmarais2024meta} that, after peer review, reports the implementation of a noncollinear version of the strongly constrained and appropriately normed (SCAN) functional \cite{sun2015strongly} in the Gaussian-basis ab-initio code package CRYSTAL~\cite{ErbaJCTC2023}.}
\sout{Mind that in the present work, we discuss the implementation and results on noncollinear magnetism and leave the discussion of SOC effects in the context of SCDFT for prospective publications.}

The article is structured as follows: In \Cref{sec:implementation}, the details of the implementation of the variational problem of SCDFT and the functionals expressions are given. In addition, we present a general magnetic scaling that could be applied to other XC functionals. In \Cref{sec:application}, we most notably show the \red{XC} spin torque of the Cr trimer. Moreover, on the example of collinear magnetic structures of MnO, we make a comparison of (A) a simple collinear, spin-polarized calculation, (B) a traditional noncollinear calculation using a collinear XC functional, and (C) the \red{XC} spin-torque calculation using the novel noncollinear XC functional. Finally, we close with a summary and propose future implications in \Cref{sec:conclusion}.

\section{Implementation}
\label{sec:implementation}

The novel \sout{spin-torque algorithm} \red{SCDFT framework including XC spin torque} is implemented into the Vienna ab-initio simulation package (VASP) \cite{kresse1993ab,kresse1994ab}, which employs the \sout{projector augmented-wave (PAW)} \red{PAW} method \cite{blochl1994projector,kresse1999ultrasoft}. The $2\times2$ spinor-formalism required for \red{XC} spin-torque calculations is presently already available to a large extent in the framework of \red{traditional} noncollinear calculations \cite{hobbs2000fully}. In addition, we reuse parts of the code programmed to accommodate meta-GGAs \cite{sun2011self}. 

In a nutshell, the implementation entails computing the $2\times 2$ paramagnetic spin-current density within the PAW method and extending the variational problem to vary the total energy with respect to \red{the} spin-current density. Naturally, this necessitates the introduction of the XC vector potential. \red{We applied special care to account for extra terms in the computation of forces and stress, as well as the fact that the paramagnetic spin-current density transforms like a rank 2 tensor with one symmetry operation in 3d Cartesian space and another in spinor space.} Finally, the expressions for the novel SCDFT functionals are shown \red{and discussed in the context of U(1)$\times$SU(2) gauge invariance}.

\subsection{Variables}

The starting point is the XC energy functional:
\begin{align}
    E_\text{xc}[&n_{\sigma\sigma'}, \mathbf{j}_{\sigma\sigma'}] =\nonumber \\ &\int\diff^3 r \, e_\text{xc}\left(n_{\sigma\sigma'},\mathbf{\nabla}n_{\sigma\sigma'},\mathbf{\nabla}^2 n_{\sigma\sigma'},\tau_{\sigma\sigma'},\mathbf{j}_{\sigma\sigma'}\right),
    \label{eq:XC dependencies}
\end{align}
which, in the present case, depends on the following $2\times2$ densities, namely the spin density
\begin{align}
    n_{\sigma \sigma'} = \sum_{i}f_i  \psi^*_{i\sigma} \psi_{i\sigma'},
    \label{eq:n}
\end{align}
where $\psi_{i\sigma}$ are the KS orbitals of the electrons and $f_i$ the corresponding occupations, the gradient $\mathbf{\nabla}n_{\sigma \sigma'}$ and Laplacian
$\mathbf{\nabla}^2n_{\sigma \sigma'}$, the kinetic-energy density
\begin{align}
    \tau_{\sigma\sigma'}&= \frac{1}{2}\sum_{i}f_i \mathbf{\nabla} \psi^*_{i\sigma} \cdot \mathbf{\nabla} \psi_{i\sigma'},
    \label{eq:tau}
\end{align}
and the paramagnetic spin-current density
\begin{align}
    \mathbf{j}_{\sigma\sigma'}  &= \frac{1}{2\imag} \sum_{i}f_i \left[\psi^*_{i\sigma}\mathbf{\nabla}\psi_{i\sigma'} - \left(\mathbf{\nabla}\psi_{i\sigma}^*\right)\psi_{i\sigma'}\right].
    \label{eq:j}
\end{align}
The expressions are given in atomic units and the sums, where $i$ is a shorthand index for the $\mathbf{k}$ point and band, run over all states.


\subsection{PAW expressions for the $2\times2$ paramagnetic spin-current density}

As for any expectation value of a semilocal operator, the paramagnetic spin-current density has the following contributions:
\begin{align}
    \mathbf{j}_{\sigma\sigma'} &= \tilde{\mathbf{j}}_{\sigma\sigma'} + \mathbf{j}^1_{\sigma\sigma'} - \tilde{\mathbf{j}}^1_{\sigma\sigma'}.
\end{align}
The \emph{valence contribution of the pseudo expectation values} on the plane-wave grid reads
\begin{align}
    \tilde{\mathbf{j}}_{\sigma\sigma'} & = \frac{1}{2\imag} \sum_{i} f_i \left[  \tilde \psi_{i\sigma}^* \mathbf{\nabla} \tilde \psi_{i\sigma'} - \left(\mathbf{\nabla} \tilde \psi_{i\sigma}^*\right)  \tilde \psi_{i\sigma'}\right],
\end{align}
where $\tilde \psi_{i\sigma}$ are the pseudo KS orbitals. 
After introducing the \emph{occupation matrix} for atomic site $R$ and the compound indices of all local quantum numbers $n$ and $n'$ that go over all augmentation channels, 
\begin{align}
    \rho_{R,nn'}^{\sigma\sigma'} &= \sum_{i} f_i \langle \tilde \psi_{i\sigma'}| \tilde p_{Rn} \rangle \langle \tilde p_{Rn'} | \tilde \psi_{i\sigma} \rangle, 
\end{align}
the \emph{valence contribution of the all-electron one-center terms} on the radial grid yields
\begin{align}
    \mathbf{j}^1_{\sigma\sigma'} &= \frac{1}{2\imag} \sum_{R,nn'} 
            \rho_{R,nn'}^{\sigma\sigma'} \left[  \phi_{Rn}^*  \mathbf{\nabla}  \phi_{Rn'} - \left( \mathbf{\nabla}  \phi_{Rn}^* \right) \phi_{Rn'}\right],
\end{align}
where $\phi_{Rn}$ are the all-electron partial waves that restore the nodal features of the KS orbitals within the PAW method. 
Likewise, the \emph{valence contribution of the pseudo one-center terms} on the radial grid is
\begin{align}
    \tilde{\mathbf{j}}^1_{\sigma\sigma'} &= \frac{1}{2\imag} \sum_{R,nn'} 
           \rho_{R,nn'}^{\sigma\sigma'} \left[  \tilde \phi_{Rn}^*  \mathbf{\nabla}   \tilde \phi_{Rn'} -  \left( \mathbf{\nabla} \tilde \phi_{Rn}^*\right)    \tilde \phi_{Rn'}\right],
\end{align}
where the pseudo partial waves $\tilde \phi_{Rn}$ are identical to $\phi_{Rn}$ outside a certain PAW radius on site $R$. Thus, the all-electron and pseudo one-center terms cancel everywhere except inside a region surrounding the atoms.
The core states are nonmagnetic and chosen as real, so that there is no contribution to the paramagnetic spin-current density. \red{See \Cref{app:Symmetrization of densities} for comments on the symmetrization of the densities.}

\subsection{Variational problem}

With the given dependencies for the XC functional in \cref{eq:XC dependencies}, the ground-state energy can be determined by applying the variational principle to the total energy \cite{Levy1982,Lieb1983,vignale1988current}

\begin{align}
\label{eq:e}
    E[n_{\sigma\sigma'},\mathbf{j}_{\sigma\sigma'}] = \min_{n_{\sigma\sigma'},\mathbf{j}_{\sigma\sigma'}} \left\{ F[n_{\sigma\sigma'},\mathbf{j}_{\sigma\sigma'}] 
    +\int \diff^3 r\, n v_\text{ext}  \right\},
\end{align}
where $v_\text{ext}$ is the external potential, due to the nuclei and the frozen core states, and
\begin{align}
\label{eq:f}
    F[n_{\sigma\sigma'},\mathbf{j}_{\sigma\sigma'}] = T_\text{KS}[n_{\sigma\sigma'},\mathbf{j}_{\sigma\sigma'}] + E_{\text{H}}[n] + E_\text{xc}[n_{\sigma\sigma'},\mathbf{j}_{\sigma\sigma'}]
\end{align}
is the universal functional, where $T_\text{KS}$ is the KS kinetic energy and $E_{\text{H}}$ is the Hartree energy. In \cref{eq:e,eq:f}, $n = n_{\uparrow\uparrow} + n_{\downarrow\downarrow}$ is the total density. Essentially, the minimization of the total energy is achieved by applying the Hamilton operator to the KS orbitals, which entails the application of the XC potential operator $\hat v_\text{xc}$ onto the KS orbitals. This corresponds to a variation of the XC energy with respect to the complex-conjugate of the KS orbitals. Hence, the following terms are computed in practice: 

\begin{widetext}
\begin{align}
    \hat{v}_{\text{xc},\sigma}\psi_{i\sigma} & = \frac{\delta E_\text{xc}}{\delta \psi^*_{i\sigma}}= \sum_{\sigma'}\left[\left( v_{\sigma\sigma'}^{\text{xc,loc}} + \mathbf{\nabla}\cdot\frac{\mathbf{A}_{\sigma\sigma'}^{\text{xc}}}{2\imag} \right) \psi_{i\sigma'} 
    - \mathbf{\nabla}\cdot\left( \frac{\mu_{\sigma\sigma'}^{\text{xc}}}{2} \mathbf{\nabla} \psi_{i\sigma'} \right) + \frac{\mathbf{A}_{\sigma\sigma'}^{\text{xc}}}{\imag} \cdot \mathbf{\nabla}  \psi_{i\sigma'}\right],
    \label{eq:var energy w.r.t. psi*}
\end{align}

\end{widetext}
where the local component reads
\begin{align}
v_{\sigma\sigma'}^{\text{xc,loc}}&=\frac{\partial e_{\text{xc}}}{\partial n_{\sigma'\sigma}} - \nabla\cdot\frac{\partial e_{\text{xc}}}{\partial\nabla n_{\sigma'\sigma}} + \nabla^2\frac{\partial e_{\text{xc}}}{\partial\nabla^2 n_{\sigma'\sigma}} ,
\label{eq:v loc}
\end{align}
the potential arising from the dependency on $\tau_{\sigma'\sigma}$ is given by
\begin{align}
\mu_{\sigma\sigma'}^{\text{xc}}&=\frac{\partial e_{\text{xc}}}{\partial\tau_{\sigma'\sigma}},
\label{eq:mu}
\end{align}
and the nonabelian \red{XC} vector potential is
\begin{align}
A_{\sigma\sigma',\alpha}^{\text{xc}} &= \frac{\partial e_{\text{xc}}}{\partial j_{\sigma'\sigma,\alpha}}.
\label{eq:A}
\end{align}
Here, $\alpha=x$, $y$, or $z$ is the Cartesian direction.

Note that the part of the first term and the entire last term which contain $\mathbf{A}_{\sigma\sigma'}^{\text{xc}}$ in \cref{eq:var energy w.r.t. psi*} are unique to SCDFT \cite{vignale1988current}.
$v^{\text{xc,loc}}_{\sigma\sigma'}$ can be identified to the \red{XC} potential for pure density functionals, i.e., LSDA, GGA, and the deorbitalized MGGA functionals, e.g., SCAN-L \cite{Mejia2017}. The term in \cref{eq:var energy w.r.t. psi*} involving $\mu_{\sigma\sigma'}^{\text{xc}}$ is present in usual MGGA functionals, e.g., R2SCAN \cite{furness2020accurate}.

\subsection{Expressions for the exchange-correlation functional}
\label{subsec:Expression for the exchange-correlation functional}

Here, we present the detailed expressions for the noncollinear exchange and correlation functionals used in this work. They are based on the
Becke-Roussel exchange functional \cite{becke1989exchange} and the Colle-Salvetti \cite{colle1975approximate} correlation functional. \blue{The expressions for the functional derivatives can be found in \Cref{sec:potbr89,sec:potcs}.}

\subsubsection{\red{Laplacian-free noncollinear} Becke-Roussel exchange functional (\red{LFNC}BR89)}
\label{subsubsec:BR89}

Following the derivation by Pittalis \emph{et al.} \cite{pittalis2017u} yields the exchange-energy functional:
\begin{align}
\label{eq:ex}
 E_{\text{x}} = 2\pi \int \diff^3 r \, n(\mathbf{r}) \int \diff s \, s\, h_{\text{x}}(\mathbf{r},s),
\end{align}
where
\begin{align}
 h_{\text{x}}(\mathbf{r},s) = - \left[ n_\text{top}(\mathbf{r}) + Q^{\red{\gamma=1}}_{\text{x}}(\mathbf{r}) s^2 + \mathcal{O}(s^4)\right]
 \label{eq:hx}
\end{align}
is the spherical average of the exchange-hole function
\begin{align}
    h_\text{x}(\mathbf{r},\mathbf{r}')=-\frac{\sum_{\sigma\sigma'}\sum_if_i\psi_{i\sigma}^*(\mathbf{r})\psi_{i\sigma'}^*(\mathbf{r}')\sum_jf_j\psi_{j\sigma'}^*(\mathbf{r}')\psi_{j\sigma}^*(\mathbf{r})}{n(\mathbf{r})}.
\end{align}
This is expanded in a Taylor series at small electron-electron distance $s$, neglecting contributions from the fourth and higher orders in $s$.
The on-top exchange hole at $s=0$ reads
\begin{align}
    n_\text{top}&= \sum_{\sigma\sigma'} \frac{n_{\sigma\sigma'}n_{\sigma'\sigma}}{n}, \label{eq:ntop}
\end{align}
and the exchange-hole curvature reads \footnote{$Q_{\text{x}}$ with $\gamma=1$ corresponds to Eq. (27) in Ref.~\cite{pittalis2017u}.}
\begin{align}
    Q_{\text{x}} &=\frac{1}{6} \left[ L -4 \gamma \left(
    \bar{\tau}  - \tau_\text{W}^\text{ncl}\right) \right].
    \label{eq:Q}
\end{align}
Here, we define the noncollinear versions of the Laplacian term
\begin{align}
    L &= \frac{\sum_{\sigma\sigma'}  n_{\sigma\sigma'} \mathbf{\nabla}^2 n_{\sigma'\sigma} }{n}, \label{eq:L definition}
\end{align}
the nonncollinear von Weizsäcker kinetic-energy density \footnote{The evaluation of the gradient should be understood as follows:
$
    \tau_\text{W}^\text{ncl}(\mathbf{r}) = \frac{\sum_{\sigma\sigma'} \left.\mathbf{\nabla}' n_{\sigma\sigma'}(\mathbf{r}')\right|_{\mathbf{r}'=\mathbf{r}}\cdot\left.\mathbf{\nabla}'' n_{\sigma'\sigma}(\mathbf{r}'')\right|_{\mathbf{r}''=\mathbf{r}}  }{8 n}
$}
\begin{align}
    \tau_\text{W}^\text{ncl} &= \frac{\sum_{\sigma\sigma'} \mathbf{\nabla} n_{\sigma\sigma'}\cdot \mathbf{\nabla} n_{\sigma'\sigma}}{8 n},
    \label{eq:tauwncl}
\end{align}
and the kinetic-energy density including the paramagnetic spin-current density
\begin{align}
\label{eq:taubar}
    \bar \tau &= \frac{\sum_{\sigma\sigma'} n_{\sigma\sigma'}\tau_{\sigma'\sigma} }{n}  - \frac{\sum_{\sigma\sigma'} \mathbf{j}_{\sigma\sigma'}\cdot\mathbf{j}_{\sigma'\sigma}  }{2n}.
\end{align}
To fulfill the \red{nonmagnetic} homogeneous electron-gas limit, the parameter $\gamma$ in \cref{eq:Q} was found to be $\gamma=0.8$ \cite{becke1989exchange}, \red{while the U(1)$\times$SU(2) gauge invariant exchange-hole curvature in \cref{eq:hx} is obtained with $\gamma=1$}.

Alternatively to \cref{eq:Q}, the exchange-hole curvature can be given by 
\begin{align}
    \tilde{Q}_{\text{x}}=\frac{1}{6} \left[ -8 \tau_\text{W}^\text{ncl}  -4 \gamma \left( 
    \bar \tau - \tau_\text{W}^\text{ncl}\right) \right]
    \label{eq:Q tilde}
\end{align}
after integrating \cref{eq:ex} by part to get rid of the Laplacian term. \red{This integration by part changes the total energy and potential, and thus the Laplacian-free noncollinear Becke-Roussel exchange
functional is not expected to yield the same results, e.g., for relaxing the volume, as the formulation using \cref{eq:Q}}. This is numerically favorable and indeed used in the implementation. One could even define a version fully independent of $\tau_\text{W}^\text{ncl}$, which yields
\begin{align}
    \bar Q_\text{x} = \frac{1}{6}\left[L-4\gamma \left(\bar\tau + \frac{1}{8}L\right)\right].\label{eq:Q x bar}
\end{align}
We will refer to this equation in the discussion of the correlation energy.

Then, based on \cref{eq:ntop} and either \cref{eq:Q}, \cref{eq:Q tilde} or \cref{eq:Q x bar} the usual Becke-Roussel expressions \cite{becke1989exchange} can be employed, leading to
\begin{align}
    E_{\text{x}} = \frac{1}{2} \int\diff^3 r  \, n(\mathbf{r}) U(\mathbf{r}),
    \label{eq:Ex}
\end{align}
where
\begin{align}
    U = -2(\pi n_{\text{top}})^{1/3}e^{x/3}\frac{1-e^{-x}\left(1+\frac{1}{2}x\right)}{x},
    \label{eq:U}
\end{align}
where $x$ is obtained by solving the following nonlinear one-dimensional equation that has no simple algebraic solution:
\begin{align}
    \frac{x e^{-2x/3}}{x-2} &= \frac{2}{3} \pi^{2/3} \frac{n_{\text{top}}^{5/3}}{\tilde{Q}_{\text{x}}}.
    \label{eq:x}
\end{align}

To inspect the nonmagnetic limit\red{,} it is useful to rewrite the above quantities in spinor space in the basis spanned by the $2\times2$ unit matrix and the Pauli matrices $\sigma_a$ with $a=1,2,3$. This way, the relevant quantities, defined in \cref{eq:ntop,eq:tauwncl,eq:taubar}, are given in terms of their charge and magnetization contributions:

\begin{align}
    n_\text{top} &= \frac{n}{2} \left(1+\frac{\sum_{a} m_a m_a}{n^2}\right), \label{eq:ntop 2}\\
    \label{eq:tauwnclnm} \tau_\mathrm{W}^\mathrm{ncl} &= \frac{\tau_\mathrm{W}}{2}\left(1+\frac{\sum_{a} \left\vert\mathbf{\nabla} m_a\right\vert^2}{\left\vert\mathbf{\nabla} n\right\vert^2}\right), 
\end{align}
where
\begin{align}
    \tau_\text{W}=\frac{|\mathbf{\nabla}n|^2}{8n} \label{eq:tauW}
\end{align}
and
\begin{align}
\label{eq:taunm}
    \bar{\tau} = \frac{\tau}{2}\left( 1 + \frac{\sum_a m_a \tau_{m,a}}{n\tau} \right) 
    - \frac{|\mathbf{j}|^2}{4n} \left( 1 + \frac{\sum_a \left\vert\mathbf{J}_a\right\vert^2}{|\mathbf{j}|^2} \right).
\end{align}
Mind that the magnetization $\vec{m}$ is a spinor quantity and not a vector-like quantity such as the charge current $\mathbf{j}$. Thus, the magnetic current $\vec{\mathbf{J}}$ is a tensor connecting spinor and vectorial quantities, which in ordinary SDFT only happens via SOC.
Based on \cref{eq:ntop 2,eq:tauwnclnm,eq:taunm}, the nonmagnetic limits are deduced:
\begin{align}
    \lim_{m_a\to 0} n_{\text{top}} &= \frac{n}{2},\\
    \lim_{m_a\to 0} \tau_\text{W}^\text{ncl} &= \frac{\tau_\text{W}}{2},\\
    \lim_{m_a\to 0} \bar \tau &= \frac{1}{2}\left(\tau - \frac{|\mathbf{j}|^2}{2n}\right),
\end{align}
which leads to the following nonmagnetic limit for $\tilde Q_{\text{x}}$:
\begin{equation}
    \lim_{m_a\to 0} \tilde Q_{\text{x}} = \frac{1}{12}\left[-8  \tau_\text{W} - 4 \gamma\left( \tau - \frac{|\mathbf{j}|^2}{2n} - \tau_\text{W} \right)\right].
\end{equation}

In our implementation, we use $\tilde Q_{\text{x}}$ as defined in \cref{eq:Q tilde}, because the second derivatives of the density, required to compute the Laplacian, introduce significant noise in the XC potential. Nevertheless, for the sake of completeness, the Laplacian in charge and magnetization representation and the corresponding nonmagnetic limit are 
\begin{equation}
    L = \frac{\mathbf{\nabla}^2 n}{2}\left( 1 + \frac{\sum_a m_a \mathbf{\nabla}^2 m_a}{n\mathbf{\nabla}^2 n} \right), \label{eq:L definition 2}
\end{equation}
\begin{equation}
    \lim_{m_a\to 0} L = \frac{\mathbf{\nabla}^2 n}{2},
\end{equation}
and
\begin{equation}
    \lim_{m_a\to 0} Q_{\text{x}} = \frac{1}{12}\left[\mathbf{\nabla}^2 n - 4 \gamma\left( \tau - \frac{|\mathbf{j}|^2}{2n} - \tau_\text{W} \right)\right].
\end{equation}

\subsubsection{\red{Noncollinear} Colle-Salvetti correlation functional (\red{NC}CS)}
\label{subsubsec:CS}

Based on the following ansatz for the many-body wavefunction by Colle and Salvetti \cite{colle1975approximate}
\begin{align}
    \Psi(\mathbf{r}_1\sigma_1,\ldots,& \mathbf{r}_N\sigma_N) \nonumber \\ &= \Psi_{\text{HF}}(\mathbf{r}_1\sigma_1,\ldots, \mathbf{r}_N\sigma_N) \prod_{i>j} \left[1-\varphi(\mathbf{r}_i,\mathbf{r}_j)\right],
\end{align}
and the definitions of the Hartree-Fock one and two-electron reduced density matrices \cite{mcweeny1960some} 
\begin{widetext}
\begin{align}
    \rho^{\sigma_1\sigma_1'}_{1\text{HF}}(\mathbf{r}_1,\mathbf{r}'_1) =& N \sum_{\sigma_2,\ldots,\sigma_N}\int \diff^3r_2\ldots\diff^3r_N\Psi_\text{HF}(\mathbf{r}_1\sigma_1,\mathbf{r}_2\sigma_2,\ldots, \mathbf{r}_N\sigma_N)\Psi_\text{HF}^*(\mathbf{r}'_1\sigma'_1,\mathbf{r}_2\sigma_2, \ldots, \mathbf{r}_N\sigma_N),\\
    \rho^{\sigma_1\sigma_2\sigma_1'\sigma_2'}_{2\text{HF}}(\mathbf{r}_1, \mathbf{r}_2;\mathbf{r}'_1,\mathbf{r}'_2) =& N(N-1) \sum_{\sigma_3,\ldots,\sigma_N}\int \diff^3r_3\ldots\diff^3r_N\Psi_\text{HF}(\mathbf{r}_1\sigma_1,\mathbf{r}_2\sigma_2,\mathbf{r}_3\sigma_3,\ldots, \mathbf{r}_N\sigma_N)\nonumber\\
    &\times\Psi_\text{HF}^*(\mathbf{r}'_1\sigma'_1,\mathbf{r}_2'\sigma'_2,\mathbf{r}_3\sigma_3,\ldots, \mathbf{r}_N\sigma_N),
\end{align}
where $\rho_{1\text{HF}}(\mathbf{r},\mathbf{r})=n(\mathbf{r})$,
Colle and Salvetti \cite{colle1975approximate} formulate an approximate form for the correlation energy:
\begin{align}
    E_\text{c}=-2a\int\diff^3 r \frac{\rho_{2\text{HF}}(\mathbf{r},\mathbf{r};\mathbf{r},\mathbf{r})}{n(\mathbf{r})} \frac{1+bn(\mathbf{r})^{-8/3}\left.\mathbf{\nabla}^2_\mathbf{s} \rho_{2\text{HF}}(\mathbf{r}-\frac{\mathbf{s}}{2},\mathbf{r}+\frac{\mathbf{s}}{2};\mathbf{r}-\frac{\mathbf{s}}{2},\mathbf{r}+\frac{\mathbf{s}}{2})\right|_{\mathbf{s}=0} e^{-cn(\mathbf{r})^{-1/3}}}{1+dn(\mathbf{r})^{-1/3}},
\end{align}
where the parameters $a=0.04918$, $b=0.06598$, $c=0.25328$, and $d=0.34935$ were fitted to the exact solution for the He atom at various distances by Colle and Salvetti \cite{colle1975approximate}.

A nonmagnetic density functional can be derived using the approximation proposed by Lee, Yang, and Parr \cite{Lee1988lyp}
\begin{align}
    \rho_{2\text{HF}}(\mathbf{r}_1,\mathbf{r}_2;\mathbf{r}_1,\mathbf{r}_2)= n(\mathbf{r}_1)n(\mathbf{r}_2)-\frac{1}{2}\rho_{1\text{HF}}(\mathbf{r}_1,\mathbf{r}_2)\rho_{1\text{HF}}(\mathbf{r}_2,\mathbf{r}_1).
\end{align}
\red{The relevance of including the orbital current density to preserve U(1) gauge invariance in a collinear framework in the Colle-Salvetti correlation energy has been demonstrated for atomic orbitals \cite{pittalis2007orbital}.} For the noncollinear version, we use the same approximation as proposed by
Tancogne-Dejean, Rubio, and Ullrich \cite{tancogne2023constructing,tancogne2023constructing_erratum}
\begin{align}
    \rho_{2\text{HF}}(\mathbf{r}_1,\mathbf{r}_2;\mathbf{r}_1,\mathbf{r}_2)= n(\mathbf{r}_1)n(\mathbf{r}_2)-\sum_{\sigma_1\sigma_2}\rho^{\sigma_1\sigma_2}_{1\text{HF}}(\mathbf{r}_1,\mathbf{r}_2)\rho^{\sigma_2\sigma_1}_{1\text{HF}}(\mathbf{r}_2,\mathbf{r}_1), \label{eq:CS rho2HF ncl}
\end{align}
which consists of the Hartree contribution minus an exchange contribution.

In agreement with Tancogne-Dejean, Rubio, and Ullrich \cite{tancogne2023constructing,tancogne2023constructing_erratum}, we find that \footnote{Mind the factor $1/2$ in the definition of the kinetic-energy density and in the two-electron reduced density matrix.}
\begin{align}
    \rho_{2\text{HF}}(\mathbf{r},\mathbf{r};\mathbf{r},\mathbf{r})= 
    n^2 - nn_{\text{top}},
    \label{eq:CS on top ncl}
\end{align}
where $n_\mathrm{top}$ is defined in \cref{eq:ntop,eq:ntop 2}, and
\begin{align}
    \left.\mathbf{\nabla}^2_\mathbf{s} \rho_{2\text{HF}}\left(\mathbf{r}-\frac{\mathbf{s}}{2},\mathbf{r}+\frac{\mathbf{s}}{2};\mathbf{r}-\frac{\mathbf{s}}{2},\mathbf{r}+\frac{\mathbf{s}}{2}\right)\right|_{\mathbf{s}=0} = 
    \frac{n\mathbf{\nabla}^2n}{2} - 4n\tau_\text{W} - \left( \frac{nL}{2}- 4n\bar\tau\right),\label{eq:CS curvature}
\end{align}
where $L$ is defined in \cref{eq:L definition,eq:L definition 2}, $\bar \tau$ in \cref{eq:taubar,eq:taunm} and $\tau_\text{W}$ in \cref{eq:tauW}.
Here, we can identify the exchange-hole curvature
\begin{align}
    \frac{nL}{2}-4n\bar \tau=6n\bar Q_\text{x}^{\gamma=1}
    \label{eq:identify Q in CS}
\end{align} 
by comparison of \cref{eq:CS curvature,eq:Q x bar} with $\gamma=1$. 
Finally, this yields the expression for the correlation energy (same as Eq.~(14) in Ref.~\cite{tancogne2023constructing}): 
\begin{align}
    E_\text{c} = -2a\int \diff^3r (n-n_{\mathrm{top}})\frac{1+bn^{-5/3} \left( \frac{\mathbf{\nabla}^2n}{2} - 4\tau_\text{W} - 6\bar Q_\text{x}^{\gamma=1}\right) e^{-cn^{-1/3}}}{1+dn^{-1/3}}. \label{eq:CS energy}
\end{align}
\end{widetext}
\sout{In fact, integration by parts can be exploited to replace $\bar Q_\text{x}^{\gamma=1}$ with $Q_\text{x}^{\gamma=1}$ or $\tilde Q_\text{x}^{\gamma=1}$ that are defined in 
This is not obvious from the final form of $E_\text{c}$ in 
but follows from the derivation of Colle and Salvetti [55].}

In the nonmagnetic limit, we obtain
\begin{align}
    \lim_{m^a\to0} & (n-n_{\mathrm{top}}) = \frac{n}{2} \label{eq:nonmag CS 1} \\
    \lim_{m^a\to0} &\left( \frac{\mathbf{\nabla}^2n}{2} -  4\tau_\text{W} - 6\bar Q_\text{x}^{\gamma=1}\right) \nonumber \\
    &\quad = \frac{\mathbf{\nabla}^2n}{4}-4\tau_\text{W} + 2 \left(\tau - \frac{|\mathbf{j}|^2}{2n}\right). \label{eq:nonmag CS 2}
\end{align}

Inserting \cref{eq:nonmag CS 1,eq:nonmag CS 2} in \cref{eq:CS energy} recovers the nonmagnetic case discussed by Lee, Yang, and Parr, see Eq.~(10) of Ref.~\cite{Lee1988lyp}.

\subsection{Application to other exchange-correlation functionals}
\label{subsec:Application to other exchange-correlation functionals}

In the course of analyzing the nonmagnetic limit in the previous sections, we could see in \cref{eq:ntop 2,eq:tauwnclnm,eq:taunm,eq:L definition 2} that each quantity has a general magnetic scaling behavior with respect to the nonmagnetic limit. This is distinct from the spin-scaling relation that is usually used to construct spin-polarized semi-local XC functionals \cite{Oliver1979scaling}. 
In order to generalize other XC functionals to the noncollinear case, one has to identify parts that originate from the on-top exchange hole and the exchange curvature. Although the substitution of $n$ and $\tau$ suggested by Pittalis \emph{et al.} \cite{pittalis2017u}, would be very elegant, generally in practice one ought to consider if the contribution is due to an exchange process and only apply magnetic scaling to those terms. For instance, in the Colle-Salvetti correlation functional discussed in \cref{eq:CS rho2HF ncl} the second contribution on the right side corresponds to an exchange process. In addition, one should carefully consider U(1)$\times$SU(2) symmetry \red{as discussed in the following section}.

\red{\section{Considerations on U(1)$\times$SU(2) gauge invariance}}

Let us recall that under a U(1)$\times$SU(2) gauge transformation the KS orbitals transform as 
\begin{align}
    \psi_{i\sigma}(\mathbf{r}) &\to \psi'_{i\sigma}(\mathbf{r}) = \sum_{\sigma'}\text{e}^{\imag\chi(\mathbf{r})}U_{\sigma\sigma'}(\mathbf{r}) \psi_{i\sigma'}(\mathbf{r}),\label{eq:gauge trafo}\\
    U_{\sigma\sigma'}(\mathbf{r}) &= \text{e}^{\imag\sum_a \lambda^a(\mathbf{r})\sigma^a_{\sigma\sigma'}},
\end{align}
where the U(1) parameter $\chi(\mathbf{r})$ and the SU(2) parameters $\lambda^a(\mathbf{r})$ are arbitrary real-valued functions in real space. Additionally, the covariant derivative should transform covariantly
\begin{align}
    \sum_{\sigma'}\mathbf{D}_{\sigma\sigma'}\psi_{i\sigma'} &\to  \sum_{\sigma'\sigma''}\text{e}^{\imag\chi(\mathbf{r})}U_{\sigma\sigma'}\left(\mathbf{D}_{\sigma'\sigma''}\psi_{i\sigma''}\right),\\
    \mathbf{D}_{\sigma\sigma'} &= \mathbf{\nabla}\delta_{\sigma\sigma'} + i\mathbf{A}_{\sigma\sigma'}.
\end{align}
Finally, the XC energy must be invariant under a U(1)$\times$SU(2) gauge transformation
\begin{align}
    E_{\text{xc}}[n,m_a,\mathbf{j},\mathbf{J}_a ]= E_{\text{xc}}[n',m'_a,\mathbf{j}',\mathbf{J}'_a ].
\end{align}
We now summarize how to check if $E_{\text{xc}}$ and individual terms therein are gauge invariant:

\begin{itemize}
    \item Insert the U(1)$\times$SU(2) transformation given in \cref{eq:gauge trafo} into the definitions of the densities \cref{eq:n,eq:tau,eq:j} to obtain the general transformation laws for SCDFT.

    \item Identify combinations of densities that appear in individual terms in $E_{\text{xc}}$ specific to the XC functional.
    \item Apply the transformation laws to the individual terms in $E_{\text{xc}}$ and confirm their U(1)$\times$SU(2) gauge invariance.
    \item Deduce the general transformation of the XC potentials for SCDFT using the transformation laws of the density, the invariance of $E_{\text{xc}}$ and the transformation of the covariant derivative.
    \item Demonstrate the covariance of the full KS operator $\hat{v}_{\text{xc},\sigma}$, c.f.~\cref{eq:var energy w.r.t. psi*}.
\end{itemize}
In \Cref{app:Transformation laws of the densities,app:Terms in the exchange-correlation energy,app:Transformation laws of XC potentials and XC spin torque}, we apply the above procedure to the XC functional presented in \cref{subsubsec:BR89} and \cref{subsubsec:CS} and show that $\gamma\neq1$ is unsuitable for U(1)$\times$SU(2) gauge-invariant calculations. We also see that in the nonmagnetic limit, the contributing terms reduce to $\tau_\text{W}$ and $\tau - \mathbf{j}^2/(2n) - \tau_\text{W}$, or rather $(\tau - \mathbf{j}^2/(2n) + \tau_\text{W})$ for $\gamma=1$. For a nonmagnetic calculation, one may only consider the U(1) transformation and find $\tau_\text{W}$ and $\tau - \mathbf{j}^2/(2n)$ to be U(1) gauge invariant, but not $\tau$ alone. Hence, for systems with finite orbital angular momentum, the inclusion of $\tau - \mathbf{j}^2/(2n)$ instead of $\tau$ alone should be preferred. This may concern e.g. f-electron systems with unquenched orbital angular momentum, $+U$ calculations including an on-site Coulomb interaction, computing the chemical shielding \cite{Dobson1993} or systems that observe finite effects due to SOC.

\section{Applications}
\label{sec:application}

We selected two systems to illustrate the particular characteristics of the first SCDFT functional including spin-torque effects implemented in the VASP code: An isolated Cr$_3$ molecule and bulk antiferromagnetic MnO, that are noncollinear and collinear magnetic systems, respectively.

\subsection{Cr$_3$ molecule}
\label{subsec:Isolated Cr3}

\begin{table}[b] 
\caption{\label{tab1:magnetic moment}%
On-site magnetic moment $\mu$ in $\mub$ of an isolated Cr$_3$ molecule for different XC functionals representative of the various families of functionals.
}
\begin{ruledtabular}
\begin{tabular}{ll}
\textrm{XC functional}&
$\mathbf{\mu}$ ($\mub$) \\
\colrule
SCDFT-\red{LFNC}BR89-\red{NC}CS & \red{$3.11(0)$}\\
MGGA-\red{LFC}BR89-\red{C}CS ($\vec{m}\times\vec{B}_{\text{xc}}\equiv0$, $\mathbf{j}_{\sigma\sigma'}\equiv0$) & \red{$3.11(7)$}\\
LSDA-PZ & $1.68$\\
GGA-PBE & $2.15$\\
MGGA-R2SCAN & $2.95$ \\
Hybrid-HSE06 & $3.32$\\
EXX-KLI & $3.81$\footnote{Reference \cite{tancogne2023constructing}}\\
Hartree-Fock &  $4.07$
\end{tabular}
\end{ruledtabular}
\end{table}

The Cr trimer is a common example to illustrate noncollinear magnetization \cite{kubler1988density, hobbs2000fully} and the presence of \red{XC} spin torque \cite{sharma2007first,tancogne2023constructing,tancogne2023constructing_erratum}. Here, we focus on the \red{XC} spin torque and on-site magnetic moment computed by means of the novel SCDFT-\red{LFNC}BR89-\red{NC}CS functional in comparison to other widely used XC functionals.

\begin{figure}[b]
\includegraphics[width=0.9\columnwidth]{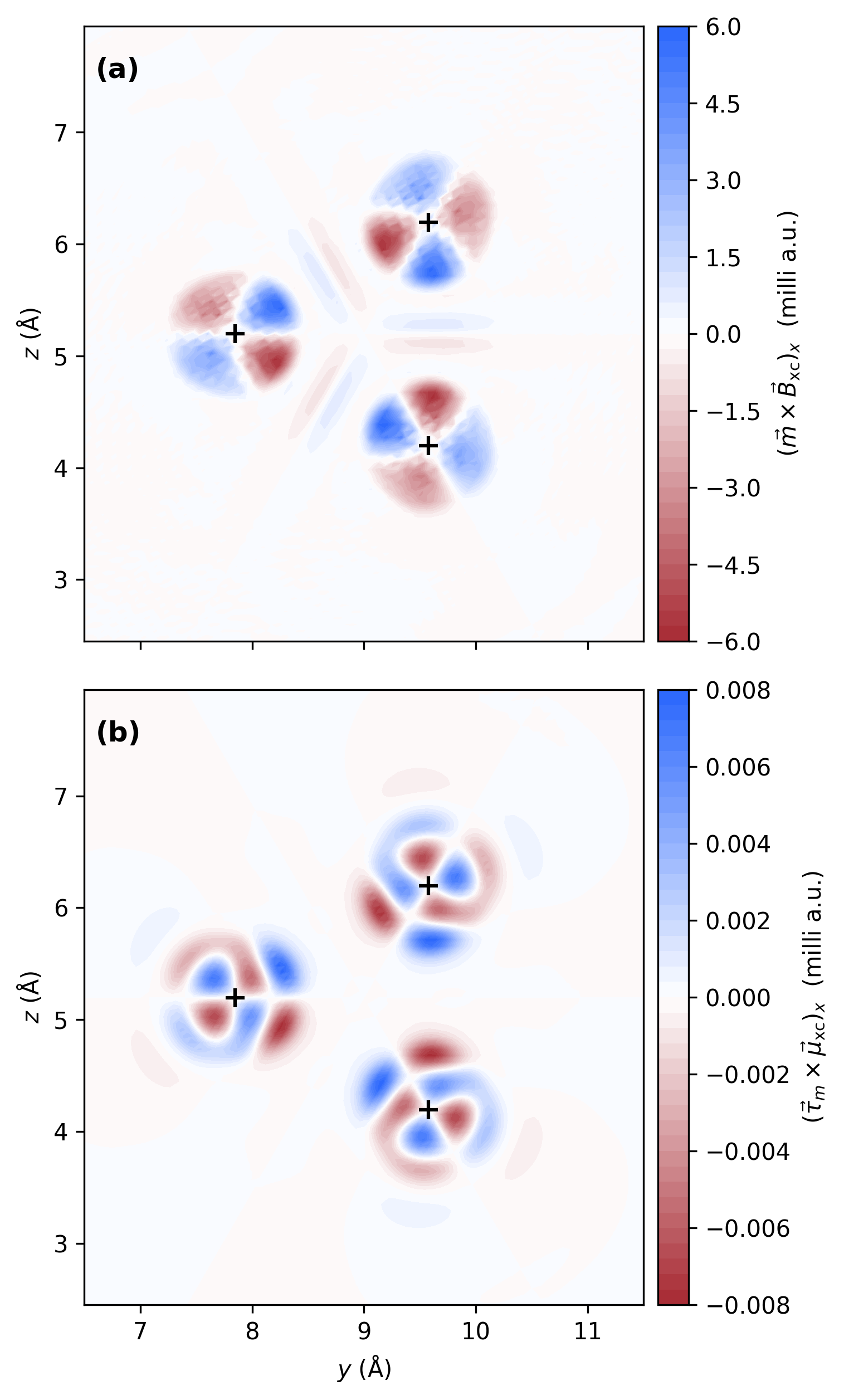}
\caption{Magnitude of \red{(a)} the \red{XC} spin torque $\vec{m} \times \vec{B}_{\text{xc}}$\red{, and (b) the kinetic \red{XC} spin torque $\vec{\tau}_m \times \vec{\mu}_{\text{xc}}$} in $10^3$~a.u.\ along the out-of-plane direction for an isolated Cr trimer using the semilocal spin-current-density-functional-theory (SCDFT) version of \red{the Laplacian-free} \blue{noncollinear} Becke-Roussel 89 exchange combined with the \red{noncollinear} Colle-Salvetti correlation, SCDFT-\red{LFNC}BR89-\red{NC}CS implemented in VASP.\label{fig2:Cr trimer spt}}
\end{figure}

The computational details of the calculations on Cr$_{3}$---presented in \cref{fig1:Cr trimer,fig2:Cr trimer spt,tab1:magnetic moment}---are as follows: The molecule is placed in a hexagonal cell as a monolayer, where the interlayer distance, as well as the distance between the centers of neighboring trimers is 8~\AA. The inter-atomic distance is $2$~\AA. The structural parameters are chosen such that the results can be compared to those in Ref.~\cite{tancogne2023constructing,tancogne2023constructing_erratum}. The calculations are performed at the $\Gamma$ point and with a cutoff energy of $35.28$~Ry for the plane-wave basis. The magnetic moments on the Cr atoms are initialized to point outwards and remain in that orientation after electronic relaxation. The magnetization and the XC B-field in \cref{fig1:Cr trimer,fig2:Cr trimer spt} are shown at every $6$th grid point in real space and we zoom into the relevant region with high electron density, i.e., we omit showing the vacuum region. We report that the magnetization and XC B-field for LSDA and SCDFT-\red{LFNC}BR89-\red{NC}CS are well behaved in the vacuum region and show no noise or unphysical values.

Remarkably, in \cref{fig2:Cr trimer spt} we can recognize similar patterns in the \red{XC} spin torque of SCDFT-\red{LFNC}BR89-\red{NC}CS as in the more involved calculations shown in Fig.~1(a), for Slater, and (b), for EXX-KLI, of Ref.~\cite{tancogne2023constructing,tancogne2023constructing_erratum}.
In contrast to the noncollinear XC functional (also based on BR89 and CS) implemented in Ref.~\cite{tancogne2023constructing,tancogne2023constructing_erratum} that is shown in Fig.~1~(c) therein, we see a smooth behavior of the \red{XC} spin torque on the Cr sites. Our approach to constructing the SCDFT-\red{LFNC}BR89-\red{NC}CS functional is different from the approach reported in Ref.\ \cite{tancogne2023constructing,tancogne2023constructing_erratum} in the following key aspects: Namely, (i) our expression for the SCDFT-BR89 exchange energy substitutes the exchange-hole curvature $Q$ rather than the kinetic energy $\tau$ to avoid breaking U(1)$\times$SU(2) invariance, (ii) we performed an integration by part to avoid the presence of the Laplacian of the density in the SCDFT-BR89 exchange energy, and (iii) the SCDFT variational problem also optimizes with respect to the three components of the $2\times2$ paramagnetic spin-current density. Thus, both the XC functional and the variational problem are distinct from Ref.~\cite{tancogne2023constructing,tancogne2023constructing_erratum}.

Strikingly, in \cref{fig2:Cr trimer spt}, four distinct regions of a larger magnitude can be observed at each site, which shows a resemblance to the EXX-KLI result of Ref.~\cite{tancogne2023constructing}. Although no symmetry is imposed during the spin-torque calculation, the total \red{XC} spin torque vanishes. Therefore, the zero-torque theorem \red{as proposed }by Capelle \emph{et al.}~\cite{capelle2001spin} \red{ and extended to SCDFT by Desmarais \emph{et al.}~\cite{Desmarais2026}} is fulfilled. \red{We see that the XC spin torque, shown in \cref{fig2:Cr trimer spt} (a), is dominant compared to the kinetic XC spin torque, in \cref{fig2:Cr trimer spt} (b).} The sign changes in the same pattern, with adjacent regions on the same site observing the opposite sign. The sign of the spin torque of our calculations agrees with the sign of Slater and EXX-KLI results in the center of the three Cr atoms, but curiously, on-site, the sign is the opposite. In between the center of the molecule and the on-site regions, there is a very thin stripe where the sign changes. Thus, it does not appear to be the on-site region that extends toward the center of the molecule but a more involved pattern similar to what is observed with the Slater potential \cite{tancogne2023constructing}. We speculate that different XC functionals within SCDFT may yield different patterns, yet the common features among the SCDFT-\red{LFNC}BR89-\red{NC}CS, EXX and EXX-KLI calculations are already compelling.  

\Cref{tab1:magnetic moment} lists the on-site spin magnetic moment $\mu$ extracted by projection onto the one-center PAW functions. In addition to the values obtained with SCDFT-\red{LFNC}BR89-\red{NC}CS, the results obtained from various other XC functionals representative for each family of functionals are also shown. 
These are (i) the LSDA Slater exchange \cite{dirac1930note} combined with Perdew-Zunger parametrization \cite{perdew1981self} of Ceperley-Alder Monte Carlo correlation data for the homogeneous electron gas \cite{ceperley1980ground} (LSDA-PZ), (ii) the Perdew-Burke-Ernzerhof GGA (GGA-PBE) \cite{perdew1996generalized} functional, (iii) the second regularized version of the strongly constrained and appropriately normed functional \cite{furness2020accurate} (MGGA-R2SCAN), (iv) the Heyd-Scuseria-Ernzerhof screened hybrid functional \cite{heyd2003hybrid,krukau2006influence} (HSE06), and finally (v) the Hartree-Fock approximation.

The results show that SCDFT-\red{LFNC}BR89-\red{NC}CS leads to a moment, \red{$\mu=3.11\mub$}, that is relatively close to the values obtained with MGGA-R2SCAN ($\mu=2.95\mub$) and HSE06 ($\mu=3.32\mub$). They are much larger than the values obtained with LSDA-PZ ($\mu=1.68\mub$) and GGA-PBE ($\mu=2.15\mub$). It is rather common that functionals of the meta-GGA and hybrid families increase the magnetic moment with respect to the standard LSDA-PZ and GGA-PBE, see e.g., Refs.~\cite{Marsman_2008,Fu2019,Tran2020magnetism} for calculations on simple ferromagnetic metals and antiferromagnetic transition-metal monoxides. At the other extreme, the Hartree-Fock approximation leads to a much larger moment of $4.07\mub$. This is rather similar to $3.81\mub$ obtained by the related EXX-KLI method \cite{tancogne2023constructing}. Mind that generally, the value of the on-site magnetic moment is sensitive to the exact definition and implementation of the on-site magnetic moment in each code. However, as both LSDA-PZ and Hartree-Fock in Ref.~\cite{tancogne2023constructing} closely agree with our values, we may point out that Tancogne-Dejean, Rubio, and Ullrich obtained a magnetic moment of $2.45\mub$, see Erratum \cite{tancogne2023constructing_erratum}, with their BR89-CS-based noncollinear functional. A value that is significantly smaller than our SCDFT-\red{LFNC}BR89-\red{NC}CS value.

As a side remark, we may mention here that hybrid functionals implicitly include spin-torque effects via the nonlocal Fock potential. However, noncollinear calculations using hybrid functionals are still problematic because the semilocal part is based on a collinear spin-polarized GGA. Recall that any not entirely local contribution to the XC energy, such as the gradient, leads to a finite spin torque, which has already been recognized by Kübler \emph{et al.} \cite{kubler1988local}.

By setting the off-diagonal elements $(\sigma,\sigma')=(\uparrow,\downarrow)$ and $(\downarrow,\uparrow)$ of the $2\times2$ densities ($n_{\sigma,\sigma'}$, $\tau_{\sigma,\sigma'}$, etc.) to zero, we obtain a collinear version of the \red{LFNC}BR89-\red{NC}CS functional. In other words, we define a collinear spin-polarized XC functional via the magnetic scaling described in \cref{subsec:Application to other exchange-correlation functionals}. This is fundamentally distinct from using the spin-scaling relation \cite{Oliver1979scaling} to obtain a collinear XC functional. Yet, it allows performing a noncollinear calculation the traditional way, i.e., by rotating the XC B-field to align it with the noncollinear magnetization self-consistently. This enforces the \red{XC} spin-torque to vanish everywhere in space, $\vec{m}\times\vec{B}_{\text{xc}}\equiv0$ \red{and $\vec{\tau}_m\times\vec{\mu}_{\text{xc}}\equiv0$}. Moreover, setting also the current density to zero, $\mathbf{j}_{\sigma\sigma'}\equiv0$, leads to a MGGA which we refer to as MGGA-\red{LFC}BR89-\red{C}CS in \cref{tab1:magnetic moment}. 

In \cref{tab1:magnetic moment}, we see that the MGGA-\red{LFC}BR89-\red{C}CS magnetic moment is the same as for SCDFT-\red{LFNC}BR89-\red{NC}CS. More dramatic is the change in the total energy: The total energy of MGGA-\red{LFC}BR89-\red{C}CS is lower than SCDFT-\red{LFNC}BR89-\red{NC}CS by about \red{$0.04$~eV}. This large difference is unlikely due to the small change in the on-site magnetic moment. Setting only $\mathbf{j}_{\sigma\sigma'}\equiv0$ affects neither the total energy nor the size of the magnetic moment in case of Cr$_3$, so we omit listing enforcing only $\vec{m}\times\vec{B}_{\text{xc}}\equiv0$ or only $\mathbf{j}_{\sigma\sigma'}\equiv0$ separately in \Cref{tab1:magnetic moment}. Hence, the energy difference of  \red{$0.04$~eV} is due to the spin-density being constrained to a subspace, where the underlying physics obeys SCDFT.

The expected magnitude of the spin torque may be scaled by $a=\mu_\text{SCDFT}/\mu_\text{EXX-KLI}\approx 0.82$, considering that the magnetic moment of SCDFT-\red{LFNC}BR89-\red{NC}CS is smaller than the magnetic moment in EXX-KLI, see~\cref{tab1:magnetic moment}. In practice, we find $b=\max{(\vec{m}\times\vec{B}_{\text{xc}})_\text{SCDFT}}/\max{(\vec{m}\times\vec{B}_{\text{xc}})_\text{EXX}}\approx 0.75$. Hence, SCDFT-\red{LFNC}BR89-\red{NC}CS captures roughly $b/a\approx90\%$ of the \red{XC} spin torque present in EXX-KLI for Cr$_3$.

In summary, the pattern of the spin torque of SCDFT-\red{LFNC}BR89-\red{NC}CS shows four on-site regions similar to EXX-KLI from Ref.~\cite{tancogne2023constructing}. However, the detailed shape and sign of the regions depend on the XC functional. The magnitude of the on-site magnetic moment is close to the values of MGGA-R2SCAN and hybrid-HSE06. Relative to that, SCDFT-\red{LFNC}BR89-\red{NC}CS yields about $90\%$ of the spin torque as expected based on EXX-KLI. The total energy changes significantly for the same XC functional if $\vec{m}(\mathbf{r})\,||\, \vec{B}_{\text{xc}}(\mathbf{r})$ is imposed ad hoc. In other words, taking the zero-local-spin-torque limit of an MGGA functional is a radical approximation.

\subsection{Rocksalt antiferromagnetic MnO}

\begin{figure*}
    \includegraphics[width=\textwidth]{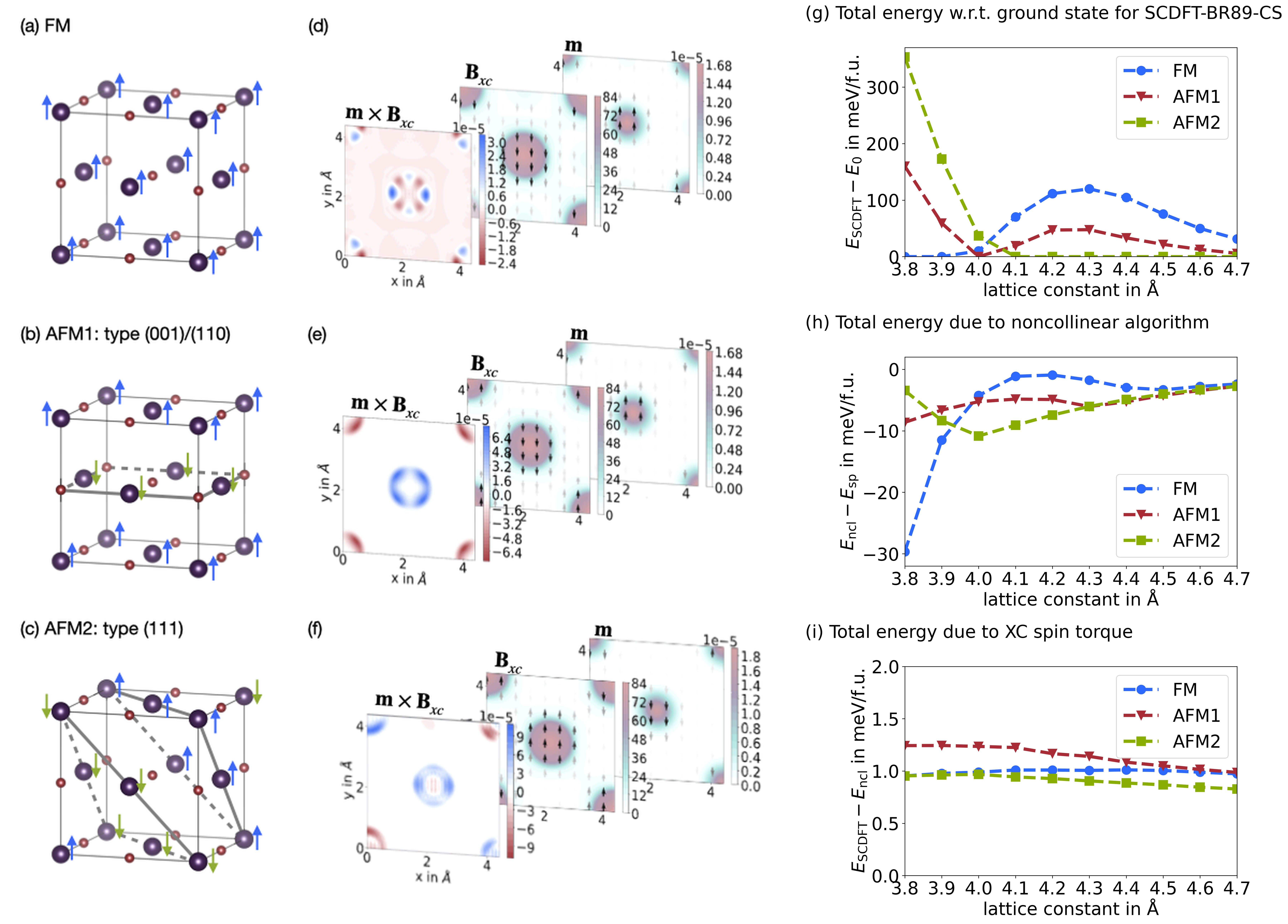}
    \caption{Magnetic configurations of MnO: (a) FM, (b) AFM1 [type (001)/(110)], and (c) AFM2 [type (111)]. (d)-(f) \red{XC} spin torque, XC B-field, and magnetization of the FM, AFM1, and AFM2 configurations, respectively, obtained from SCDFT-\red{LFNC}BR89-\red{NC}CS. (g) Total energy per formula unit of SCDFT-\red{LFNC}BR89-\red{NC}CS with respect to the ground state at the given lattice constant $a$ as a function of $a$. (h) Total energy difference between the spin-polarized and the noncollinear algorithm using the spin-polarized MGGA-\red{LFC}BR89-\red{C}CS functional. (i) Total energy difference between the spin-polarized and the noncollinear versions of the BR89-CS functional within the noncollinear algorithm. \label{fig3:MnO structure}}
\end{figure*}

\begin{table*}
\caption{\label{tab2:MnO}Properties of MnO calculated with various XC functionals using either (A) the collinear algorithm, (B) the traditional noncollinear algorithm \red{based on collinear XC functionals} neglecting \red{XC} spin torque, or (C) noncollinear algorithm within SCDFT including \red{XC} spin torque. The columns correspond to the equilibrium lattice constant $a_0$, the equilibrium lattice volume $V_0$, the bulk modulus $B_0$, \red{the fundamental band gap $E_g$,} and the on-site magnetic moment $\mu$. \red{All quantities refer to the AFM2 ground state.}}
\begin{ruledtabular}
\begin{tabular}{llllllll}
Case &
XC functional &
Algorithm   &
$a_0$ (\AA) &
$V_0$ (\AA$^3$)&
$B_0$ (GPa)&
\red{$E_g$ (eV)}&
$\mu$ ($\mub$)\\
\colrule
1 &LSDA-\red{PZ}         & (A) collinear      & 4.32 & 80 & 155 & \red{0.48} & 4.14\\
2 &MGGA-R2SCAN & (A) collinear      & 4.42 & 86 & 172 & \red{1.72} & 4.36 \\
3 &MGGA-R2SCAN & (B) noncollinear   & 4.42 & 86 & 172 & \red{1.8} & 4.36 \\
4 &MGGA-\red{LFC}BR89-\red{C}CS   & (A) collinear      & 4.33 & 81 & 191 & \red{1.16} & \red{4.36} \\
5 &MGGA-\red{LFC}BR89-\red{C}CS   & (B) noncollinear   & 4.33 & 81 & 191 & \red{1.19} & \red{4.36} \\
6 &SCDFT-\red{LFNC}BR89-\red{NC}CS  & (C) \red{XC} spin-torque    & 4.33 & 81 & 192 & \red{1.20} & \red{4.36} \\
7 &Hybrid-HSE06   & (A) collinear      & 4.44 & 87 & 164 & \red{2.98} & 4.48\\
Experiment & &                         & 4.43\footnote{Reference \cite{sasaki1980estimation}} & 87\footnote{Reference \cite{sasaki1980estimation}} & 151\footnote{Reference \cite{noguchi1996shock}}, 162\footnote{Reference \cite{jeanloz1987static}} & \red{$3.9\pm 0.4$\footnote{\red{Reference \cite{fujimori1990electronic,van1991electronic}}}} & 4.54\footnote{Reference \cite{jauch2003electron} }
\end{tabular}
\end{ruledtabular}
\end{table*}

The discussion of the new SCDFT-\red{LFNC}BR89-\red{NC}CS functional based on MnO is motivated by two aspects: (i) Archer, \emph{et al.} \cite{archer2011exchange} show analogous results for a range of XC functionals including hybrid-HSE06 and GGA-PBE, and (ii) the possibility to apply all three available modes for a magnetic calculation: 
\begin{itemize}
    \item[(A)] A collinear spin-polarized calculation using both a collinear spin-polarized XC functional and algorithm.
    \item[(B)] A traditional noncollinear calculation using a collinear spin-polarized XC functional with a noncollinear algorithm.
    \item[(C)] A \red{XC} spin-torque calculation using a noncollinear XC functional with a noncollinear algorithm.
\end{itemize}

The considered ferromagnetic (FM) and two different antiferromagnetic (AFM) \blue{\sout{magnetic}} configurations are shown in \cref{fig3:MnO structure}(a)-(c). Naively, one would expect that such collinear systems with negligible SOC are well-described with a simple collinear calculation corresponding to (A) which would make any differences compared to (B) and/or (C) rather intriguing.

The computational details for the calculations presented in \cref{fig3:MnO structure} and \Cref{tab2:MnO} are as follows: The FM and AFM1 structures consist of 8 atomic sites in the conventional unit cell, while the cell of AFM2 is chosen by considering the Mn sites on the diagonal in \cref{fig3:MnO structure}(c), i.e., the lower left and upper right Mn sites as well as the central O site and the O site in extension of the diagonal that is not shown in the figure. The lattice constant $a$ is varied from 3.8 to 4.7~\AA. In combination with these structures, the k-point mesh for the integration into the Brillouin zone is $8\times8\times8$ for FM and AFM1, and $4\times4\times4$ for AFM2. The calculations are performed at a cutoff energy of 38.22 Ry for the plane-wave basis. The magnetic moments are initialized as shown in \cref{fig3:MnO structure}(a)-(c), and no symmetrization is applied.

\Cref{fig3:MnO structure}(d)-(f) displays the characteristic patterns of the \red{predominant XC} spin torque corresponding to the magnetic configurations shown in \cref{fig3:MnO structure}(a)-(c), respectively. The \red{XC} spin torque is exceedingly small as it is five orders of magnitude smaller than in Cr$_3$. This indicates that the zero-spin-torque limit is indeed a good approximation, which is discussed in more detail below. Despite the small magnitude, the patterns capture the symmetry properties of the system and display characteristic sign changes depending on the magnetic order.

\Cref{fig3:MnO structure}(g) presents the total energy of case (C) with the SCDFT-\red{LFNC}BR89-\red{NC}CS functional for all three magnetic configurations as a function of the lattice constant $a$ relative to the total energy $E_0$ of the most stable magnetic configuration at each value of $a$. This can be directly compared to the overview of XC functionals in Fig. 4 of Ref.~\cite{archer2011exchange}. There are two phase transitions in close proximity around 4~\AA, with a narrow region where AFM1 is the most stable magnetic structure. This is qualitatively similar to what is obtained with MGGA-R2SCAN (not shown), while in contrast, hybrid-HSE06 yields no phase transition, with AFM2 being the most stable magnetic structure throughout the considered range for $a$ \cite{archer2011exchange}. The pure density functionals, LSDA and GGA-PBE, yield one phase transition from FM in the high-pressure region to AFM2 under ambient pressure.

As mentioned, the magnetic transition as a function of pressure obtained by SCDFT-\red{LFNC}BR89-\red{NC}CS closely resembles that of MGGA-R2SCAN. Furthermore, the collinear version of BR89-CS yields qualitatively the same behavior with both algorithms, the collinear spin-polarized and the traditional noncollinear algorithm. This again hints towards the zero-torque limit being a good approximation for MnO. Nevertheless, we want to closely inspect the differences in the total energy of cases (A) $E_\text{sp}$, (B) $E_\text{ncl}$, and (C) $E_\text{SCDFT}$. Surprisingly, there are relevant energy differences between the three modes as elucidated in \cref{fig3:MnO structure}(h) and (i). The fact that the magnetic ground state as a function of the lattice constant does not depend on the mode is due to the large energy difference between the magnetic structures of roughly 50-300 meV/f.u.. In other words, we cannot generally expect that the magnetic ground state is the same for (A), (B), and (C) for other systems as elaborated below.

\Cref{fig3:MnO structure}(h) presents the total energy difference $E_\text{ncl}-E_\text{sp}$ between the traditional noncollinear and spin-polarized calculations, both using the collinear spin-polarized version of BR89-CS. We see that for all investigated magnetic configurations and lattice constants the traditional noncollinear calculations lead to lower total energies. This can trivially be understood in terms of constrained minimization: The collinear algorithm (A) can only access a subset of solutions of the noncollinear case (B). In other words, the noncollinear algorithm (B) has more degrees of freedom to minimize the total energy and should strictly yield equal or lower total energy compared to the collinear algorithm (A) within the same XC functional. 

We note that the energy difference $E_\text{ncl}-E_\text{sp}$ in this system varies between approx. $-28$ meV/f.u. and $-0.8$ meV/f.u., which is surprisingly large for a system where we expected this difference to vanish. It strongly depends on the magnetic configuration and the applied pressure. Hence, the relative stability of magnetic states changes, and in systems with almost degenerate magnetic configurations, this could change the predicted magnetic ground state. Moreover, in systems with itinerant electron magnetism with noncollinear magnetization in the interstitial regions the difference between the collinear algorithm and noncollinear algorithm is expected to be much larger. This is already seen in the present data: The FM solution at lattice constant $a\leq3.9$~\AA~is metallic, while it is insulating for larger values of $a$. The metallic systems show a large total energy difference despite the fact that the magnetization is FM.

Next, let us compare (B) the traditional noncollinear algorithm using MGGA-\red{LFC}BR89-\red{C}CS and (C) the noncollinear SCDFT-\red{LFNC}BR89-\red{NC}CS. Considering the data in \cref{fig3:MnO structure}~(i), we find that the total energy obtained with SCDFT-\red{LFNC}BR89-\red{NC}CS is strictly higher than in the traditional noncollinear MGGA-\red{LFC}BR89-\red{C}CS calculation. We may understand this in terms of constrained minimization, since indeed, the spin-torque algorithm constrains the system to a physical subset of spin-densities with a certain spin-current density and spin torque.
Importantly, we see that the energy difference $E_\text{SCDFT}-E_\text{ncl}$ depends on the magnetic configuration. Hence, including spin-torque effects changes the relative stability of magnetic configurations and thus potentially changes the magnetic ground state. We also see that compared to Cr$_3$, where the energy difference is 0.05~eV, the effect is 50 times smaller, but not five orders of magnitude smaller than one may expect based on the spin torque.

Comparing \cref{fig3:MnO structure}~(h) and (i), $E_\text{SCDFT}-E_\text{ncl}$ is an order of magnitude smaller than $E_\text{ncl}-E_\text{sp}$. Hence, the spin-torque algorithm still yields equal or smaller total energies compared to the collinear algorithm. We also note that the spin-torque algorithm using the SCDFT-\red{LFNC}BR89-\red{NC}CS functional converges in fewer electronic minimization steps than the noncollinear algorithm. In fact, the spin-torque algorithm is as stable as the collinear algorithm, while the noncollinear algorithm takes about 3 times as many electronic steps for MnO. \sout{The instability of the noncollinear algorithm is likely due to the ad hoc zero local spin-torque approximation which leads to inconsistencies between the energy and the potential. The same inconsistencies are present for the collinear algorithm, however it has much fewer degrees of freedom and is thus less prone to convergence issues.}

Turning now to the results in \Cref{tab2:MnO}, we see that the lattice constant $a_0$, bulk modulus $B_0$, \red{bandgap} and on-site magnetic moment $\mu$ are sensitive to the choice of the XC functional, as expected, but almost independent of the choice of the algorithm (A)-(C). 
Similar to LSDA, the SCDFT-\red{LFNC}BR89-\red{NC}CS and MGGA-\red{LFC}BR89-\red{C}CS functional significantly underestimates the equilibrium lattice constant $a_0$ for MnO by more than $3\%$, while MGGA-R2SCAN and hybrid-HSE06 yield very good predictions for $a_0$ with less than $0.3\%$ of error. In addition, hybrid-HSE06 leads to a very good estimate for the bulk modulus $B_0$ and the on-site magnetic moment. The magnetic moments are quite accurately estimated by the \red{LF(N)C}BR89-\red{NC}CS functional, as well. \red{The influence of the framework goes beyond the accuracy of measurement in both, the calculation and experiment, e.g., due to ambiguities in the definition of the on-site moment.}

In summary, the results for MnO show that most of the physics in collinear magnetic systems without SOC is already captured by a collinear algorithm and the effects of the choice of the XC functional is much larger. The \red{LFNC}BR89-\red{NC}CS functional fails to \red{accurately }describe structural parameters, however it is the only XC functional available \red{at the time of writing} in the framework of SCDFT\red{; in a parallel work the so-called noncollinear modefied SCAN was developed \cite{desmarais2024meta}}. Nevertheless, the SCDFT-\red{LFNC}BR89-\red{NC}CS functional yields a characteristic behavior in the \red{XC} spin torque and it converges exceptionally well.

\section{Conclusion}
\label{sec:conclusion}

We have shown the first ever consistent noncollinear ab-initio calculation using a U(1)$\times$SU(2) gauge-invariant noncollinear semilocal XC functional, with a variational problem that is extended
by the variation with respect to the components of the 2×2 paramagnetic spin-current density. 
The novel XC functional is formulated within SCDFT and cures the lack of \red{XC} spin torque in semilocal XC functionals. For the Cr$_3$ molecule we have shown that the \red{XC} spin torque is quite similar to the one obtained by EXX-KLI \cite{tancogne2023constructing}.

Our \red{XC} spin-torque calculations fundamentally differ from traditional noncollinear calculations \cite{kubler1988density,kubler1988local} since the latter are based on collinear XC functionals. These traditional noncollinear calculations introduce an ad hoc zero local \red{XC} spin-torque approximation and often seem to suffer under poor\sout{which introduces an inconsistency between XC energy and XC potential that may lead to problems for the} electronic convergence even in cases where the approximation is small. For instance, in MnO the total energy correction due to the local \red{XC} spin torque is merely of the order of 1 meV/f.u., while the \red{XC} spin-torque algorithm needs roughly a third of the number of steps in the electronic minimization compared to the traditional noncollinear algorithm. In magnetic systems with noncollinear magnetic order, the influence on the total energy due to neglecting \red{XC} spin torque can be much larger and potentially changes the magnetic ground state. For example in Cr$_3$, the \red{change in the} total energy due to spin torque is \red{40}~meV, which is well above the common energy differences between different magnetic orders \cite{huebsch2021benchmark}.

Collinear XC functionals should be derived from SCDFT by setting the off-diagonal terms of the $2\times 2$ quantities to zero instead of the commonly employed spin-scaling relation. That is because collinear systems also disobey exact spin scaling unless they are homogeneous electron gases, i.e., a pure spin-up or spin-down state. 
The expressions for our novel implementation are based on the Becke-Roussel exchange and Colle-Salvetti correlation functionals. It depends on the $2\times 2$ spin density, kinetic-energy density, and spin-current density. The implementation in VASP entails computing the paramagnetic spin-current density within the PAW method and extending the variational problem to vary the total energy with respect to the density current, which makes it computationally slightly more demanding than MGGA calculations.

While the ad hoc zero local \red{XC} spin-torque approximation strongly affects the total energy and convergence properties, the on-site magnetic moment and lattice parameters are not sensitive to it for the investigated systems. In other words, traditional noncollinear and collinear algorithms yield practically the same lattice constant, bulk modulus, and magnetic moment. This is good news because established XC functionals are promising candidates to be extended to SCDFT in order to cure the lack of \red{XC} spin torque, bad electronic convergence, and the discrepancy between the experiment and the total energy landscape of SDFT while still yielding satisfying results for other properties.

For future work, several directions are worth exploring. On the one hand, there is a plethora of magnetic phenomena that may be sensitive to the zero local \red{XC} spin-torque approximation. On the other hand, more established XC functionals should be extended to the framework of SCDFT to allow for U(1)$\times$SU(2) gauge-invariant noncollinear calculations without resorting to the ad hoc local zero \red{XC} spin-torque approximation.

\begin{acknowledgments}

\blue{
The authors are very grateful to Stefano Pittalis and Jacques Desmarais for fruitful discussions on various fundamental aspects,
in particular the U(1)$\times$SU(2) invariance.
We also thank Nicolas Tancogne-Dejean for helpful discussions regarding the implementation reported in Refs.~\cite{tancogne2023constructing,tancogne2023constructing_erratum}.
}

\end{acknowledgments}

\appendix

\section{Transformation laws of the densities}
\label{app:Transformation laws of the densities}

Inserting \cref{eq:gauge trafo} into the definitions of the $2\times2$ densities in \cref{eq:n,eq:tau,eq:j}, yields the transformation laws for the densities \cite{vignale1988current,pittalis2017u}
\begin{subequations}
\begin{align}
    n \to n'=n
\end{align}
\begin{align}
    m_a \to m'_a= \sum_b R_{ab}m_b
\end{align}
\begin{widetext}
\begin{align}
    \tau \to \tau' =\ \tau &
    + \mathbf{\nabla} \chi \cdot \mathbf{j} 
    + \frac{1}{2}\left(\mathbf{\nabla}\chi\right)^2 n 
    -\frac{\imag}{2} \sum_a \mathbf{J}_a \cdot \mathrm{Tr}\left\{\sigma_aU^\dagger\mathbf{\nabla}U\right\}
    + \frac{1}{8} \sum_a \mathrm{Tr}\left\{\sigma_aU^\dagger\mathbf{\nabla}U\right\} \cdot \mathrm{Tr}\left\{\sigma_aU^\dagger\mathbf{\nabla}U\right\}
\end{align}
\begin{align}
    \tau_{m,a} \to \tau'_{m,a} =& \sum_b R_{ab} \left[ \tau_{m,b} + \mathbf{\nabla}\chi\cdot\mathbf{J}_b + \frac{1}{2} \left(\nabla\chi\right)^2m_b 
     - \frac{\imag}{2} \mathbf{j} \cdot \mathrm{Tr}\left\{\sigma_bU^\dagger\mathbf{\nabla}U\right\}\right. \nonumber \\
     &+ \frac{1}{8} \mathrm{Tr}\left\{\sigma_bU^\dagger\mathbf{\nabla}U\right\} \cdot \sum_c \mathrm{Tr}\left\{\sigma_cU^\dagger\mathbf{\nabla}U\right\} m_c 
    \left.- \frac{1}{8} m_b \sum_c \mathrm{Tr}\left\{\sigma_cU^\dagger\mathbf{\nabla}U\right\} \cdot \mathrm{Tr}\left\{\sigma_cU^\dagger\mathbf{\nabla}U\right\}\right] \nonumber \\
    &+ \sum_b \mathbf{\nabla} R_{ab} \cdot\mathbf{\nabla} m_{b}
\end{align}
\end{widetext}
\begin{align}
    \mathbf{j}  \to \mathbf{j}' = \mathbf{j} + (\mathbf{\nabla} \chi) n - \frac{\imag}{2} \sum_am_a\mathrm{Tr}\left\{\sigma_aU^\dagger\mathbf{\nabla}U\right\}
\end{align}
\begin{align}
    \mathbf{J}_a \to \mathbf{J'}_a = R_{ab} \left[ \mathbf{J}_b + (\mathbf{\nabla}\chi) m_b - \frac{\imag}{2}n\mathrm{Tr}\left\{\sigma_bU^\dagger\mathbf{\nabla}U\right\} \right]
\end{align}
\label{eq:density trafo}
\end{subequations}
where we used
\begin{align}
    \sum_b R^{ab} \sigma^b_{\sigma\sigma'} &= \sum_{\sigma'',\sigma'''}U^\dagger_{\sigma\sigma'''} \sigma^a_{\sigma'''\sigma''}U_{\sigma''\sigma'},
\end{align}
and
\begin{align}    
    \mathrm{Tr}\left\{\sigma^bU^\dagger\mathbf{\nabla}U\right\} &= \sum_{\sigma\sigma'\sigma''} \sigma_{\sigma\sigma''}^b U^\dagger_{\sigma''\sigma'} \mathbf{\nabla}U_{\sigma'\sigma}.
\end{align}
Here, $R_{ab}$ is a rotation in the $\mathbb{R}^3$ space spanned by the Pauli matrices around $\vec{\lambda}=(\lambda_1,\lambda_2,\lambda_3)^T$ by an angle of $-2|\vec{\lambda}|$.

\section{Terms in the exchange-correlation energy}
\label{app:Terms in the exchange-correlation energy}

\subsection{Laplacian-free noncollinear Becke-Roussel exchange functional (LFNCBR89)}

\subsubsection{Gauge invariance}

\Cref{eq:Ex,eq:U,eq:x} show that the following combination of densities appears as individual terms in the Laplacian-free noncollinear BR89 exchange functional: $n_{\text{top}}$ defined in \cref{eq:ntop,eq:ntop 2}, and $\tilde Q_{\text{x}}$ given in \cref{eq:Q tilde}. These comprise of $\tau_\text{W}^\text{ncl}$ and $\bar \tau$. While for $\gamma=1$ the sum $(\bar \tau +\tau_\text{W}^\text{ncl})$ enters, for $\gamma\neq1$ both $\tau_\text{W}^\text{ncl}$ and $(\bar \tau -\tau_\text{W}^\text{ncl})$ should be considered separately. 

After identifying these terms, we can apply the transformations listed in \cref{eq:density trafo} and obtain the following transformation laws:
\begin{subequations}
\begin{align}
    n_{\text{top}} &\to n_{\text{top}},\\
    \tilde Q_{\text{x}}^{\gamma=1} &\to
    \tilde Q_{\text{x}}^{\gamma=1}\\
    \bar \tau +     \tau^{\text{ncl}}_\text{W} &\to \bar \tau + \tau^{\text{ncl}}_\text{W}
\end{align}
\end{subequations}
which are all gauge invariant. 

However, considering the case of $\gamma=0.8$, or more generally $\gamma\neq1$, we find
\begin{widetext}
\begin{subequations}
\begin{align}
    \tau^{\text{ncl}}_\text{W} &\to (\tau^{\text{ncl}}_\text{W})' = \tau^{\text{\text{ncl}}}_\text{W} + \frac{1}{16n} \left( \sum_a m_a m_a + 2 \sum_{abc} m_a (\mathbf{\nabla} R_{ab}) R_{bc} (\mathbf{\nabla} m_c) \right) ,\\
    \bar \tau - \tau^{\text{ncl}}_\text{W} = \bar \tau + \tau^{\text{ncl}}_\text{W} - 2\tau^{\text{ncl}}_\text{W} &\to \bar \tau' - (\tau^{\text{ncl}}_\text{W})' = \bar \tau - \tau^{\text{ncl}}_\text{W} - \frac{1}{8n} \left( \sum_a m_a m_a + 2 \sum_{abc} m_a (\mathbf{\nabla} R_{ab}) R_{bc} (\mathbf{\nabla} m_c) \right) 
\end{align}
\label{eq: gauge breaking of 0.8}
\end{subequations}

using 
\begin{subequations}
\begin{align}
    \mathbf{\nabla} m_a &\to (\mathbf{\nabla} m_a)'= (\mathbf{\nabla} R_{ab}) m_b + R_{ab} \mathbf{\nabla} m_b,\\
    4n\bar \tau = 2n\tau + 2 \sum_a m_a \tau_{m,a} - |\mathbf{j}|^2 - \sum_a \mathbf{J}_a\cdot\mathbf{J}_a &\to 4n\bar \tau -  \frac{1}{4} \left( \sum_a m_a m_a + 2 \sum_{abc} m_a (\mathbf{\nabla} R_{ab}) R_{bc} (\mathbf{\nabla} m_c) \right).
\end{align}
\end{subequations}

\end{widetext}
As visible from \cref{eq: gauge breaking of 0.8} using $\gamma=0.8$ is unsuitable for U(1)$\times$SU(2) gauge invariant calculations. 

\subsubsection{\blue{Functional derivative}}
\label{sec:potbr89}

The expression of the functional derivative of the noncollinear BR89 functional, \cref{eq:Ex}, is provided below.
The partial derivatives of the exchange-energy density per volume $e_\text{x}=\left(1/2\right)nU$ entering into \cref{eq:var energy w.r.t. psi*,eq:v loc,eq:mu,eq:A}
are given by the following expressions:
\begin{equation}
\frac{\partial e_\text{x}}{\partial n_{\sigma\sigma'}} =
\frac{1}{2}\left(\left(\delta_{\sigma\sigma'}+\frac{1}{3}\frac{n}{n_{\text{top}}}\frac{\partial n_{\text{top}}}{\partial n_{\sigma\sigma'}}\right)U+n\frac{\partial U}{\partial x}\frac{\partial x}{\partial n_{\sigma\sigma'}}\right)
\label{eq:dexd1}
\end{equation}
\begin{equation}
\frac{\partial e_\text{x}}{\partial\left(\nabla n_{\sigma\sigma'}\cdot\nabla n_{\sigma''\sigma'''}\right)} = \frac{1}{2}n\frac{\partial U}{\partial x}\frac{\partial x}{\partial\left(\nabla n_{\sigma\sigma'}\cdot\nabla n_{\sigma''\sigma'''}\right)}
\label{eq:dexd2}
\end{equation}
\begin{equation}
\frac{\partial e_\text{x}}{\partial\tau_{\sigma\sigma'}} = \frac{1}{2}n\frac{\partial U}{\partial x}\frac{\partial x}{\partial\tau_{\sigma\sigma'}}
\label{eq:dexd3}
\end{equation}
\begin{equation}
\frac{\partial e_\text{x}}{\partial\mathbf{j}_{\sigma\sigma'}} = \frac{1}{2}n\frac{\partial U}{\partial x}\frac{\partial x}{\partial\mathbf{j}_{\sigma\sigma'}}
\label{eq:dexd4}
\end{equation}
where the various quantities appearing in \cref{eq:dexd1,eq:dexd2,eq:dexd3,eq:dexd4} are given by the following expressions:
\begin{equation}
\frac{\partial n_{\text{top}}}{\partial n_{\sigma\sigma'}} = \frac{2n_{\sigma'\sigma}-\delta_{\sigma\sigma'}n_{\text{top}}}{n}
\end{equation}
\begin{equation}
\frac{\partial U}{\partial x} =-2(\pi n_{\text{top}})^{1/3}e^{-2x/3}\frac{x^2+2x+e^x(x-3)+3}{3x^2}
\end{equation}
\begin{equation}
\frac{\partial x}{\partial n_{\sigma\sigma'}} =
\frac{(x-2)x}{x^2-2x+3}\left(\frac{3}{2}\frac{1}{\tilde{Q}_{\text{x}}}\frac{\partial \tilde{Q}_{\text{x}}}{\partial n_{\sigma\sigma'}}-
\frac{5}{2}\frac{1}{n_{\text{top}}}\frac{\partial n_{\text{top}}}{\partial n_{\sigma\sigma'}}\right)
\end{equation}
\begin{align}
\frac{\partial x}{\partial\left(\nabla n_{\sigma\sigma'}\cdot\nabla n_{\sigma''\sigma'''}\right)} =
\frac{(x-2)x}{x^2-2x+3}\frac{3}{2} \nonumber \\
\times \frac{1}{\tilde{Q}_{\text{x}}}\frac{\partial \tilde{Q}_{\text{x}}}{\partial\left(\nabla n_{\sigma\sigma'}\cdot\nabla n_{\sigma''\sigma'''}\right)}
\end{align}
\begin{equation}
\frac{\partial x}{\partial \tau_{\sigma\sigma'}} =
\frac{(x-2)x}{x^2-2x+3}\frac{3}{2}\frac{1}{\tilde{Q}_{\text{x}}}\frac{\partial \tilde{Q}_{\text{x}}}{\partial \tau_{\sigma\sigma'}}
\end{equation}
\begin{equation}
\frac{\partial x}{\partial\mathbf{j}_{\sigma\sigma'}} =
\frac{(x-2)x}{x^2-2x+3}\frac{3}{2}\frac{1}{\tilde{Q}_{\text{x}}}\frac{\partial \tilde{Q}_{\text{x}}}{\partial\mathbf{j}_{\sigma\sigma'}}
\end{equation}
where the partial derivatives of $\tilde{Q}_{\text{x}}$ [\cref{eq:Q tilde}] are given by
\begin{equation}
\frac{\partial \tilde{Q}_{\text{x}}}{\partial n_{\sigma\sigma}} = \frac{2}{3n}\left[\gamma\left(\bar{\tau}-\tau_{\sigma\sigma}\right)+(2-\gamma)\tau_{\text{W}}^{\text{ncl}}\right]
\end{equation}
\begin{equation}
\frac{\partial \tilde{Q}_{\text{x}}}{\partial n_{\uparrow\downarrow}} = -\frac{2}{3}\gamma\frac{\tau_{\downarrow\uparrow}}{n}
\end{equation}
\begin{equation}
\label{eq:dqdg1}
\frac{\partial \tilde{Q}_{\text{x}}}{\partial(\nabla n_{\sigma\sigma}\cdot\nabla n_{\sigma'\sigma'})} = \delta_{\sigma\sigma'}\frac{\gamma-2}{12}\frac{1}{n}
\end{equation}
\begin{equation}
\label{eq:dqdg2}
\frac{\partial \tilde{Q}_{\text{x}}}{\partial(\nabla n_{\uparrow\downarrow}\cdot\nabla n_{\downarrow\uparrow})} = \frac{\gamma-2}{6}\frac{1}{n}
\end{equation}
\begin{equation}
\frac{\partial \tilde{Q}_{\text{x}}}{\partial\tau_{\sigma\sigma'}} = -\frac{2}{3}\gamma\frac{n_{\sigma'\sigma}}{n}
\end{equation}
\begin{equation}
\frac{\partial \tilde{Q}_{\text{x}}}{\partial\mathbf{j}_{\sigma\sigma'}} = \frac{2}{3}\gamma\frac{\mathbf{j}_{\sigma'\sigma}}{n}
\end{equation}

\subsection{Noncollinear Colle-Salvetti correlation functional (CS)}

\subsubsection{Gauge invariance}

\Cref{eq:CS energy,eq:CS on top ncl,eq:CS curvature} show that the correlation energy is expressed in terms of $n$, $n_{\text{top}}$, $n\nabla^2n-8n\tau_\text{W}$, and $nL-8n\bar\tau$. In addition to the U(1)$\times$SU(2) gauge invariance of $n$, $n_{\text{top}}$, and $\tau_\text{W}$, as discussed in the previous section, we find

\begin{subequations}
\begin{align}
    n\nabla^2n &\to n\nabla^2n,\\
    nL-8n\bar\tau=12n\bar Q_\text{x}^{\gamma=1} &\to 12n\bar Q_\text{x}^{\gamma=1}.
\end{align}
\end{subequations}
For the latter, note that 
\begin{align}
    \frac{\nabla^2}{4n} \left(nn + \sum_{a}m_am_a\right) = 8\tau_\text{W}^\text{ncl} + L &\to 8\tau_\text{W}^\text{ncl} + L,
\end{align}
and $\bar\tau$ and $\tau_\text{W}^\text{ncl}$ transform such that $\bar\tau+\tau_\text{W}^\text{ncl}$ stays invariant.

\subsubsection{\blue{Functional derivative}}
\label{sec:potcs}

The functional derivative of the noncollinear CS functional, \cref{eq:CS energy}, is given below.
We start by rewriting \cref{eq:CS energy} as follows:
\begin{equation}
E_\text{c} = \int \diff^3 r e_\text{c} = -a\int \diff^3 r n\zeta_{\text{c}} f
\end{equation}
where $\zeta_{\text{c}}=1-|\vec{m}|^2/n^2=2\left(1-n_{\text{top}}/n\right)$ and
\begin{equation}
f = \frac{g}{h}
\end{equation}
with
\begin{equation}
g = 1+\bar{b}r_s^5ke^{-\bar{c}r_s}
\end{equation}
\begin{equation}
h = 1+\bar{d}r_s
\end{equation}
where
\begin{equation}
k = \frac{\mathbf{\nabla}^2n}{2} - 4\tau_\text{W} - 6\bar Q_\text{x}^{\gamma=1}
\end{equation}
and $\bar{b}=b\left(4\pi/3\right)^{5/3}$, $\bar{c}=c\left(4\pi/3\right)^{1/3}$,
and $\bar{d}=d\left(4\pi/3\right)^{1/3}$.
The various partial derivatives of $e_\text{c}$ are given by
\begin{equation}
\label{eq:decd1}
\frac{\partial e_\text{c}}{\partial n_{\sigma\sigma'}} = -a\left(\delta_{\sigma\sigma'}\zeta_{\text{c}} f + n\frac{\partial\zeta_{\text{c}}}{\partial n_{\sigma\sigma'}}f+n\zeta_{\text{c}}\frac{\partial f}{\partial n_{\sigma\sigma'}}\right)
\end{equation}
\begin{equation}
\label{eq:decd2}
\frac{\partial e_\text{c}}{\partial\left(\nabla n_{\sigma\sigma'}\cdot\nabla n_{\sigma''\sigma'''}\right)} =
-an\zeta_{\text{c}}\frac{\partial f}{\partial\left(\nabla n_{\sigma\sigma'}\cdot\nabla n_{\sigma''\sigma'''}\right)}
\end{equation}
\begin{equation}
\label{eq:decd3}
\frac{\partial e_\text{c}}{\partial\left(\nabla^2 n_{\sigma\sigma'}\right)} =
-an\zeta_{\text{c}}\frac{\partial f}{\partial\left(\nabla^2 n_{\sigma\sigma'}\right)}
\end{equation}
\begin{equation}
\label{eq:decd4}
\frac{\partial e_\text{c}}{\partial\tau_{\sigma\sigma'}} =
-an\zeta_{\text{c}}\frac{\partial f}{\partial\tau_{\sigma\sigma'}}
\end{equation}
\begin{equation}
\label{eq:decd5}
\frac{\partial e_\text{c}}{\partial\mathbf{j}_{\sigma\sigma'}} =
-an\zeta_{\text{c}}\frac{\partial f}{\partial\mathbf{j}_{\sigma\sigma'}}
\end{equation}
where the various quantities appearing in \cref{eq:decd1,eq:decd2,eq:decd3,eq:decd4,eq:decd5} are given by the following expressions:

\begin{equation}
\frac{\partial\zeta_{\text{c}}}{\partial n_{\sigma\sigma'}} = \frac{2}{n}\left(1-\zeta_{\text{c}}\right) \delta_{\sigma\sigma'} - 4  \frac{n_{\sigma'\sigma} }{n^2}\sigma_{\sigma\sigma'}^1 - 2\frac{n_{\sigma'\sigma}}{n^2} \sigma_{\sigma\sigma'}^3 
\end{equation}

\begin{equation}
\frac{\partial f}{\partial n_{\sigma\sigma'}} =
\frac{1}{h^2}\left(\frac{\partial g}{\partial n_{\sigma\sigma'}}h-g\frac{\partial h}{\partial n_{\sigma\sigma'}}\right)
\end{equation}

\begin{equation}
\frac{\partial f}{\partial\left(\nabla n_{\sigma\sigma'}\cdot\nabla n_{\sigma''\sigma'''}\right)} =
\frac{1}{h}\frac{\partial g}{\partial\left(\nabla n_{\sigma\sigma'}\cdot\nabla n_{\sigma''\sigma'''}\right)}
\end{equation}

\begin{equation}
\frac{\partial f}{\partial\nabla^2 n_{\sigma\sigma'}} =
\frac{1}{h}\frac{\partial g}{\partial\nabla^2 n_{\sigma\sigma'}}
\end{equation}

\begin{equation}
\frac{\partial f}{\partial\tau_{\sigma\sigma'}} =
\frac{1}{h}\frac{\partial g}{\partial\tau_{\sigma\sigma'}}
\end{equation}

\begin{equation}
\frac{\partial f}{\partial\mathbf{j}_{\sigma\sigma'}} =
\frac{1}{h}\frac{\partial g}{\partial\mathbf{j}_{\sigma\sigma'}}
\end{equation}

\begin{equation}
\frac{\partial h}{\partial n_{\sigma\sigma'}} = \bar{d}\frac{\partial r_s}{\partial n_{\sigma\sigma'}}
\end{equation}

\begin{equation}
\frac{\partial r_s}{\partial n_{\sigma\sigma'}} = -\delta_{\sigma\sigma'}\frac{1}{3}\frac{r_s}{n}
\end{equation}

\begin{equation}
\frac{\partial g}{\partial n_{\sigma\sigma'}} =\bar{b}\Bigg[
r_s^5\frac{\partial k}{\partial n_{\sigma\sigma'}} +
5r_s^4\frac{\partial r_s}{\partial n_{\sigma\sigma'}}k -
r_s^5k\bar{c}\frac{\partial r_s}{\partial n_{\sigma\sigma'}}\Bigg]e^{-\bar{c}r_s}
\end{equation}

\begin{equation}
\frac{\partial g}{\partial\left(\nabla n_{\sigma\sigma'}\cdot\nabla n_{\sigma''\sigma'''}\right)} =
\bar{b}r_s^5\frac{\partial k}{\partial\left(\nabla n_{\sigma\sigma'}\cdot\nabla n_{\sigma''\sigma'''}\right)}e^{-\bar{c}r_s}
\end{equation}

\begin{equation}
\frac{\partial g}{\partial\nabla^2 n_{\sigma\sigma'}} =
\bar{b}r_s^5\frac{\partial k}{\partial\nabla^2 n_{\sigma\sigma'}}e^{-\bar{c}r_s}
\end{equation}

\begin{equation}
\frac{\partial g}{\partial\tau_{\sigma\sigma'}} =
\bar{b}r_s^5\frac{\partial k}{\partial\tau_{\sigma\sigma'}}e^{-\bar{c}r_s}
\end{equation}

\begin{equation}
\frac{\partial g}{\partial\mathbf{j}_{\sigma\sigma'}} =
\bar{b}r_s^5\frac{\partial k}{\partial\mathbf{j}_{\sigma\sigma'}}e^{-\bar{c}r_s}
\end{equation}

\begin{equation}
\frac{\partial k}{\partial n_{\sigma\sigma'}} = 
\delta_{\sigma\sigma'}\frac{1}{n}\left(6\tilde{Q}_{\text{x}} + 4\tau_{\text{W}}\right)+
\frac{1}{n}\left(4\tau_{\sigma'\sigma} -\frac{1}{2}\nabla^2 n_{\sigma'\sigma}\right)
\end{equation}

\begin{equation}
\frac{\partial k}{\partial\left(\left\vert\nabla n_{\sigma\sigma}\right\vert^2\right)} =
-\frac{1}{2n}
\end{equation}

\begin{equation}
\frac{\partial k}{\partial\left(\nabla n_{\uparrow\uparrow}\cdot\nabla n_{\downarrow\downarrow}\right)} =
-\frac{1}{n}
\end{equation}

\begin{equation}
\frac{\partial k}{\partial(\nabla n_{\uparrow\downarrow}\cdot\nabla n_{\downarrow\uparrow})} = 0
\end{equation}

\begin{equation}
\frac{\partial k}{\partial\nabla^2 n_{\sigma\sigma'}} = \frac{1}{2}\left(\delta_{\sigma\sigma'}-\frac{n_{\sigma'\sigma}}{n}\right)
\end{equation}

\begin{equation}
\frac{\partial k}{\partial\tau_{\sigma\sigma'}} = 4\frac{n_{\sigma'\sigma}}{n}
\end{equation}

\begin{equation}
\frac{\partial k}{\partial\mathbf{j}_{\sigma\sigma'}} = -2\frac{\mathbf{j}_{\sigma\sigma'}}{n}
\end{equation}

\section{Transformation laws of XC potentials and covariance of the Kohn-Sham operator}
\label{app:Transformation laws of XC potentials and XC spin torque}

The XC energy must be invariant
\begin{align}
    E_{\text{xc}}[n,m_a,\mathbf{j},\mathbf{J}_a ]= E_{\text{xc}}[n',m'_a,\mathbf{j}',\mathbf{J}'_a ].
\end{align}
Further combining the invariance of $E_{\text{xc}}$ with the transformation of the densities yields the transformation of the XC potentials defined in \cref{eq:var energy w.r.t. psi*,eq:v loc,eq:mu,eq:A}. For a more intuitive picture, we decompose the $2\times2$ potentials into charge (scalar) and magnetization (vector) parts using $n_{\sigma\sigma'} = \frac{1}{2}(n\,\delta_{\sigma\sigma'} + \sum_a m_a\,\sigma^a_{\sigma\sigma'})$ and analogously for the kinetic energy density, the currents and all potentials.

\subparagraph{Local potential.}
Writing $v_{\sigma\sigma'}^{\mathrm{xc,loc}} = v^{\mathrm{xc,loc}}_0\,\delta_{\sigma\sigma'} + \sum_a B_{\mathrm{xc},a}\,\sigma^a_{\sigma\sigma'}$, the scalar XC potential $v^{\mathrm{xc,loc}}_0$ and the XC magnetic field $\vec{B}_{\mathrm{xc}}$ transform as:
\begin{align}
    v^{\mathrm{xc,loc}}_0 &\to \left.v^{\mathrm{xc,loc}}_0\right.'=v^{\mathrm{xc,loc}}_0 , \\
    B_{\mathrm{xc},a} &\to \left. B_{\mathrm{xc},a} \right.'=\sum_b R_{ab}\, B_{\mathrm{xc},b}.
\end{align}
That is, the scalar potential is gauge invariant and the XC $B$-field rotates with the same $R$ matrix as the magnetization.

\subparagraph{Kinetic potential.}
Defining a scalar auxiliary quantity 
\begin{align}
    \bar\mu = \left.\frac{\partial e_{\mathrm{xc}}}{\partial\bar\tau}\right|_{n_{\sigma\sigma'}}
\end{align}
and noting that
\begin{align}
    \frac{\partial\bar\tau}{\partial\tau_{\sigma'\sigma}} = \frac{n_{\sigma\sigma'}}{n},
\end{align}
we find the XC kinetic potential can be written as
\begin{align}
    \mu_{\sigma\sigma'}^{\mathrm{xc}} &= \frac{\partial e_{\mathrm{xc}}}{\partial\tau_{\sigma'\sigma}} 
    = \frac{\partial e_{\mathrm{xc}}}{\partial\bar\tau}\,\frac{\partial\bar\tau}{\partial\tau_{\sigma'\sigma}}
    = \bar\mu\,\frac{n_{\sigma\sigma'}}{n}.
\end{align}
Furthermore this reads 
\begin{align}
    \mu_{\sigma\sigma'}^{\mathrm{xc}} = \frac{\bar\mu}{2n}\left(n\delta_{\sigma\sigma'} + \sum_a m_a\,\sigma^a_{\sigma\sigma'}\right),
\end{align}
in terms of charge and magnetization contributions
and transforms as:
\begin{align}
    \mu^{\mathrm{xc}}_0 = \frac{\bar\mu}{2} &\to \left.\mu^{\mathrm{xc}}_0\right.' = \mu^{\mathrm{xc}}_0, \\
    \mu^{\mathrm{xc}}_a =\frac{\bar\mu\, m_a}{2n} &\to \left. \mu^{\mathrm{xc}}_a\right.' =\frac{\bar\mu}{2n}\sum_b R_{ab}\, m_b
\end{align}
This is derived by applying the transformation laws derived for the densities and considering that both $\bar\mu$ and $n$ are gauge invariant. Hence, the scalar part is invariant and the vector part rotates with $R$.

To see that $\bar\mu$ is gauge invariant, the XC functionals must be analyzed individually. For LFNCBR89,
\begin{align}
    \bar\mu = \frac{\partial e_\mathrm{xc}}{\partial\bar\tau}\bigg|_{n_{\sigma\sigma'},\nabla n_{\sigma\sigma'}} = \frac{\partial e_\mathrm{xc}}{\partial(\bar\tau + \tau_\mathrm{W}^{\mathrm{ncl}})}\bigg|_{n_{\sigma\sigma'},\nabla n_{\sigma\sigma'}}.
\end{align}
Both the energy density $e_\mathrm{xc}$ and the argument $\bar\tau + \tau_\mathrm{W}^{\mathrm{ncl}}$ are gauge-invariant scalars, hence their ratio (the derivative) is also gauge invariant.

Note that $\tau_\mathrm{W}^{\mathrm{ncl}}$ involves $\nabla n_{\sigma\sigma'}$, which does \emph{not} transform simply under SU(2). However, 
$\tau_\mathrm{W}^{\mathrm{ncl}}$ is held fixed when differentiating with respect to $\bar\tau$, and the combination $\bar\tau + \tau_\mathrm{W}^{\mathrm{ncl}}$ is gauge invariant regardless of how its individual summands transform.

For NCCS, $\bar \tau$ also appears only in a gauge invariant combination which allows rewriting the partial derivative defining $\bar \mu$ as
\begin{align}
    \bar\mu = \frac{\partial e_\mathrm{xc}}{\partial\bar\tau}\bigg|_{n_{\sigma\sigma'},\nabla n_{\sigma\sigma'}} = -\frac{\partial e_\mathrm{xc}}{\partial(\frac{L}{8}-\bar \tau)}\bigg|_{n_{\sigma\sigma'},\nabla^2 n_{\sigma\sigma'}}
\end{align} 
and again shows $\bar \mu$ to be gauge invariant.

\subparagraph{Vector potential.}
Similarly, we can use the auxiliary potential $\bar \mu$ to write
\begin{align}
    A_{\sigma\sigma',m}^{\mathrm{xc}} &= \frac{\partial e_\mathrm{xc}}{\partial j_{\sigma'\sigma,m}} 
    = \frac{\partial e_\mathrm{xc}}{\partial\bar\tau}\,\frac{\partial\bar\tau}{\partial j_{\sigma'\sigma,m}}
    = -\bar\mu\,\frac{j_{\sigma\sigma',m}}{n}.
\end{align}
Decomposing the current as $j_{\sigma\sigma',m} = \frac{1}{2}(j_m\,\delta_{\sigma\sigma'} + \sum_a J_{a,m}\,\sigma^a_{\sigma\sigma'})$, the XC vector potential $A_{\sigma\sigma',m}^{\mathrm{xc}}$ separates into a charge (abelian) part and a spin (nonabelian) part:
\begin{align}
    A^{\mathrm{charge}}_m &= -\frac{\bar\mu\, j_m}{2n}, &
    A_{a,m}^{\mathrm{spin}} &= -\frac{\bar\mu\, J_{a,m}}{2n}.
\end{align}
\begin{widetext}
Using the transformation laws of $\mathbf{j}$ and $\mathbf{J}_a$ from Eqs.~\eqref{eq:density trafo}, these transform as:
\begin{align}
    A^{\mathrm{charge}}_m& \to \left. A^{\mathrm{charge}}_m \right.' = A^{\mathrm{charge}}_m - \frac{\bar\mu}{2}\nabla_m\chi \nonumber  + \frac{\imag\bar\mu}{4n}\sum_a m_a\,\mathrm{Tr}\left\{\sigma_a U^\dagger\nabla_m U\right\}, \\
    A_{a,m}^{\mathrm{spin}}& \to 
    \left. A_{a,m}^{\mathrm{spin}} \right.' =\sum_b R_{ab}\left[ A_{b,m}^{\mathrm{spin}} - \frac{\bar\mu\,(\nabla_m\chi)\, m_b}{2n} + \frac{\imag\bar\mu}{4}\,\mathrm{Tr}\left\{\sigma_b U^\dagger\nabla_m U\right\}\right]. 
\end{align}

The first equation shows the abelian U(1) gauge-connection behavior (shift by $-\bar\mu\nabla_m\chi/2$) plus a SU(2) mixing term. The second equation shows that the spin vector potential rotates with $R$ plus inhomogeneous (connection) terms from both U(1) and SU(2).

\subparagraph{Covariance of the full KS operator.}
The final task is to verify explicitly that the KS operator~\cref{eq:var energy w.r.t. psi*} transforms covariantly under the combined U(1)$\times$SU(2) gauge transformation:
\begin{align}
    \frac{\delta E}{\delta\psi'^*_{i\sigma}} = e^{\imag\chi}\sum_{\sigma''}U_{\sigma\sigma''}\frac{\delta E}{\delta\psi^*_{i\sigma''}}. \label{eq:covariance goal}
\end{align}

The gradient of the transformed spinor is:
\begin{align}
    \nabla_m\psi'_{i\sigma} = e^{\imag\chi}\sum_{\sigma'}\left[U_{\sigma\sigma'}\nabla_m\psi_{i\sigma'} + (\nabla_m U_{\sigma\sigma'})\psi_{i\sigma'} + \imag(\nabla_m\chi)U_{\sigma\sigma'}\psi_{i\sigma'}\right]. \label{eq:grad psi prime}
\end{align}

For the local potential, we use $v_{\sigma\sigma'}^{\mathrm{loc}} \to (Uv^{\mathrm{loc}}U^\dagger)_{\sigma\sigma'}$:
\begin{align}
    \sum_{\sigma'}v'^{\mathrm{loc}}_{\sigma\sigma'}\psi'_{i\sigma'} = e^{\imag\chi}\sum_{\sigma'\alpha\beta\gamma}U_{\sigma\alpha}\,v^{\mathrm{loc}}_{\alpha\beta}\,U^\dagger_{\beta\sigma'}\,U_{\sigma'\gamma}\,\psi_{i\gamma} = e^{\imag\chi}\sum_{\alpha}U_{\sigma\alpha}\sum_{\beta}v^{\mathrm{loc}}_{\alpha\beta}\psi_{i\beta},
\end{align}
with $\sum_{\sigma'}U^\dagger_{\beta\sigma'}U_{\sigma'\gamma} = \delta_{\beta\gamma}$. This gives the covariant for the usual GGA potential.
\end{widetext}
Secondly, we find that the transformed kinetic potential $\mu'^{\mathrm{xc}}_{\sigma\sigma'} = (U\mu^{\mathrm{xc}}U^\dagger)_{\sigma\sigma'}$ involves:
\begin{align}
    -\sum_{\sigma'}\nabla_m\left(\frac{\mu'^{\mathrm{xc}}_{\sigma\sigma'}}{2}\nabla_m\psi'_{i\sigma'}\right).
\end{align}
Inserting~\eqref{eq:grad psi prime} and the covariant transformation of $\mu^{\mathrm{xc}}$, and collecting terms, one finds the covariant piece $e^{\imag\chi}U_{\sigma\alpha}[-\nabla_m(\mu^{\mathrm{xc}}_{\alpha\beta}\nabla_m\psi_{i\beta}/2)]$ plus cross-terms involving $\nabla_m\chi$ and $\nabla_m U$. 

Finally, these cross-terms exactly cancel 
the terms arising due to the inhomogeneous part of $A'^{\mathrm{xc}}$ and leave the
covariant transformation of the terms involving $\mathbf{A}_{\sigma\sigma'}^{\mathrm{xc}}$ of ~\cref{eq:var energy w.r.t. psi*} as the remainder. Hence, in summary \cref{eq:covariance goal} is fulfilled.

\section{Symmetrization of densities}
\label{app:Symmetrization of densities}

For each crystal-symmetry operator $\tilde R\in \tilde S$ at a given $\mathbf{k}$ in the irreducible Brillouin zone, a scalar is invariant under rotations $\tilde R$ in 3d Cartesian space. Thus, the symmetrized charge density is simply the average
\begin{align}
    \bar n(\mathbf{k}) = \frac{1}{|\tilde S|} \sum_{\tilde R\in \tilde S} n(\mathbf{k}),
\end{align}
where $|\tilde S|$ is the number of crystal-symmetry operations that leave the system invariant.

The magnetization, however, transforms according to all magnetic space-group operations $R \in S$ as a pseudo-vector, which implies considering $\det{( R)}$. This reads
\begin{align}
    \bar{m}^a(\mathbf{k}) = \frac{1}{|S|} \sum_{R\in S} \sum_b R^{ab}m^b(\mathbf{k}).
\end{align}
It is noteworthy, that (magnetic) space-group operations are position-independent, causing all the problematic gradient terms in \cref{eq:density trafo} to vanish. The kinetic energy densities $\tau$ and $\vec{\tau}_m$ hence transform identically as $n$ and $\vec{m}$.

The charge current transforms like a vector, i.e., it can be symmetrized using 
\begin{align}
    \bar{j}^i(\mathbf{k}) = \frac{1}{|\tilde S|} \sum_{\tilde R\in \tilde S} \sum_{j} \tilde R^{ij}j^j(\mathbf{k}),
\end{align}
where $j=x,y,z$ corresponds to the Cartesian directions.
Finally, the magnetic current transforms like a rank 2 tensor yielding
\begin{align}
    \bar{J}^{ia}(\mathbf{k}) = \frac{1}{|\tilde S||S|} \sum_{\tilde R, R} \sum_{j=x,y,z} \sum_{b } \tilde R^{ij}J^{jb}(\mathbf{k})(R^{-1})^{ba}.
\end{align}

\bibliography{references}

\end{document}